\documentclass[%
 reprint,
 amsmath,amssymb,
prl,
aps,
showpacs,
]{revtex4-1}

\usepackage[FIGTOPCAP, hang, nooneline]{subfigure}
\usepackage{float}

\usepackage{graphicx}
\usepackage{dcolumn}
\usepackage{bm}
\usepackage{amssymb}
\usepackage{amsfonts}
\usepackage{amsmath}
\usepackage[colorlinks,hyperindex]{hyperref}

\input{epsf}
\begin{document}

\title{Radiofrequency spectroscopy of a linear array of Bose-Einstein condensates in a magnetic lattice}

\author{ P.~Surendran$^{1}$, S.~Jose$^{1}$, Y.~Wang$^{1}$, I.~Herrera$^{1}$,  H. Hu$^1$, X. Liu$^1$, S.~Whitlock$^{2}$, R.~McLean$^{1}$,
A.~Sidorov$^{1}$  and P.~Hannaford$^{1}$}
\email[]{phannaford@swin.edu.au}
\affiliation {\makebox[\textwidth]{$^1$Centre for Quantum and Optical Science, Swinburne University of Technology, Melbourne, Australia 3122.}
 $^2$Physikalisches Institut, Universit$\ddot{a}$t Heidelberg, Im Neuenheimer Feld 226, 69120 Heidelberg, Germany}

\date{\today}

\begin{abstract}

We report site-resolved radiofrequency spectroscopy measurements of Bose-Einstein condensates of $^{87}$Rb atoms in about 100 sites of a one-dimensional 10 $\mu$m-period magnetic lattice produced by a grooved magnetic film plus bias fields.  Site-to-site variations of the trap bottom, atom temperature, condensate fraction and chemical potential indicate that the magnetic lattice is remarkably uniform, with variations in the trap bottoms of only $\pm$0.4 mG. At the lowest trap frequencies (radial and axial frequencies 1.5 kHz and 260 Hz, respectively), temperatures down to 0.16 $\mu$K are achieved in the magnetic lattice and at the smallest trap depths (50 kHz) condensate fractions up to 80\% are observed. With increasing radial trap frequency (up to 20 kHz, or aspect ratio up to $\sim$ 80) large condensate fractions persist and the highly elongated clouds approach the quasi-1D Bose gas regime. The temperature estimated from analysis of the spectra is found to increase by a factor of about five which may be due to suppression of rethermalising collisions in the quasi-1D Bose gas. Measurements for different holding times in the lattice indicate a decay of the atom number with a half-life of about 0.9 s due to three-body losses and the appearance of a high temperature ($\sim$1.5 $\mu$K) component which is attributed to atoms that have acquired energy through collisions with energetic three-body decay products.
\end{abstract}
\maketitle
\section{I. Introduction}\vspace{-0.4cm}
Magnetic lattices comprising periodic arrays of magnetic microtraps created by patterned magnetic films  \cite{Ghanbari06,Gerritsma07,Singh08,Whitlock09,Abdelrahman10,Schmied10,Leung11,jose14,Leung14,ivan14} provide a potentially powerful complementary tool to optical lattices for simulating many-body condensed matter phenomena. Such lattices offer a high degree of design flexibility. They may, in principle, be tailored to produce 2D or 1D arrays of ultracold atoms in nearly arbitrary configurations \cite{Schmied10} and with arbitrary lattice spacings that are not restricted by the optical wavelength. To date, ultracold atoms have been successfully loaded into a one-dimensional magnetic lattice with period 10 $\mu$m \cite{Singh08,jose14}, a two-dimensional rectangular magnetic lattice with period about 25 $\mu$m \cite{Gerritsma07,Whitlock09,Leung11} and square and triangular magnetic lattices with period 10 $\mu$m [9]. Two-dimensional square and triangular magnetic lattices with periods down to 0.7 $\mu$m, designed for quantum tunnelling experiments, have recently been fabricated and characterised \cite{ivan14}.

In a recent Rapid Communication \cite{jose14} we reported preliminary results for the realization of a periodic array of about 100 Bose-Einstein condensates (BECs) of $^{87}$Rb $|F = 1, m_F=-1>$ atoms in a one-dimensional 10 $\mu$m-period magnetic lattice. Clear signatures for the onset of Bose-Einstein condensation were provided by $in$-$situ$ site-resolved radiofrequency (RF) spectroscopy. The atoms were prepared in the $|F=1, m_F= -1 >$ low field-seeking state which has a three-times smaller three-body recombination coefficient \cite{Burt97,Soeding99} and a weaker confinement compared with atoms in the $|F=2,m_F= +2 >$ state.

In this paper we report site-resolved radiofrequency (RF) measurements of Bose-Einstein condensates of $^{87}$Rb $|F=1, m_F= -1 >$ atoms in multiple sites of the one-dimensional 10 $\mu$m-period magnetic lattice. Radiofrequency spectra are presented for a range of trap depths, trap frequencies and holding times in the magnetic lattice, including  for highly elongated magnetic lattice traps in which the atom clouds approach the quasi-1D Bose gas regime \cite{Kinoshita06,mazets10,Jacqmin11}. The effect of magnetic noise on the RF spectra, which was not included in our earlier analysis \cite{jose14}, is found to have a significant effect on the temperature, condensate fraction, chemical potential and atom number derived from the RF spectra. 

In Section II we present background theory on the 1D magnetic lattice and a self-consistent Hartree-Fock mean-field model used to analyze the RF spectra. Section III gives experimental details including a description of the magnetic lattice atom chip, the procedure for atom cooling and loading atoms into the lattice, the trap frequency measurements and the site-resolved RF spectroscopy setup. In Section IV we present RF spectra for multiple BECs in the magnetic lattice for a range of trap depths, trap frequencies and holding times in the lattice. In Section V we summarize our results and discuss some future directions.\vspace{-0.4cm}
\section{II.	BACKGROUND THEORY}\vspace{-0.4cm}
{\bf{A. One-dimensional magnetic lattice}}\vspace{0.2cm}

\begin{figure}[htbp]
	\begin{center}
		\includegraphics[width=0.5\textwidth]{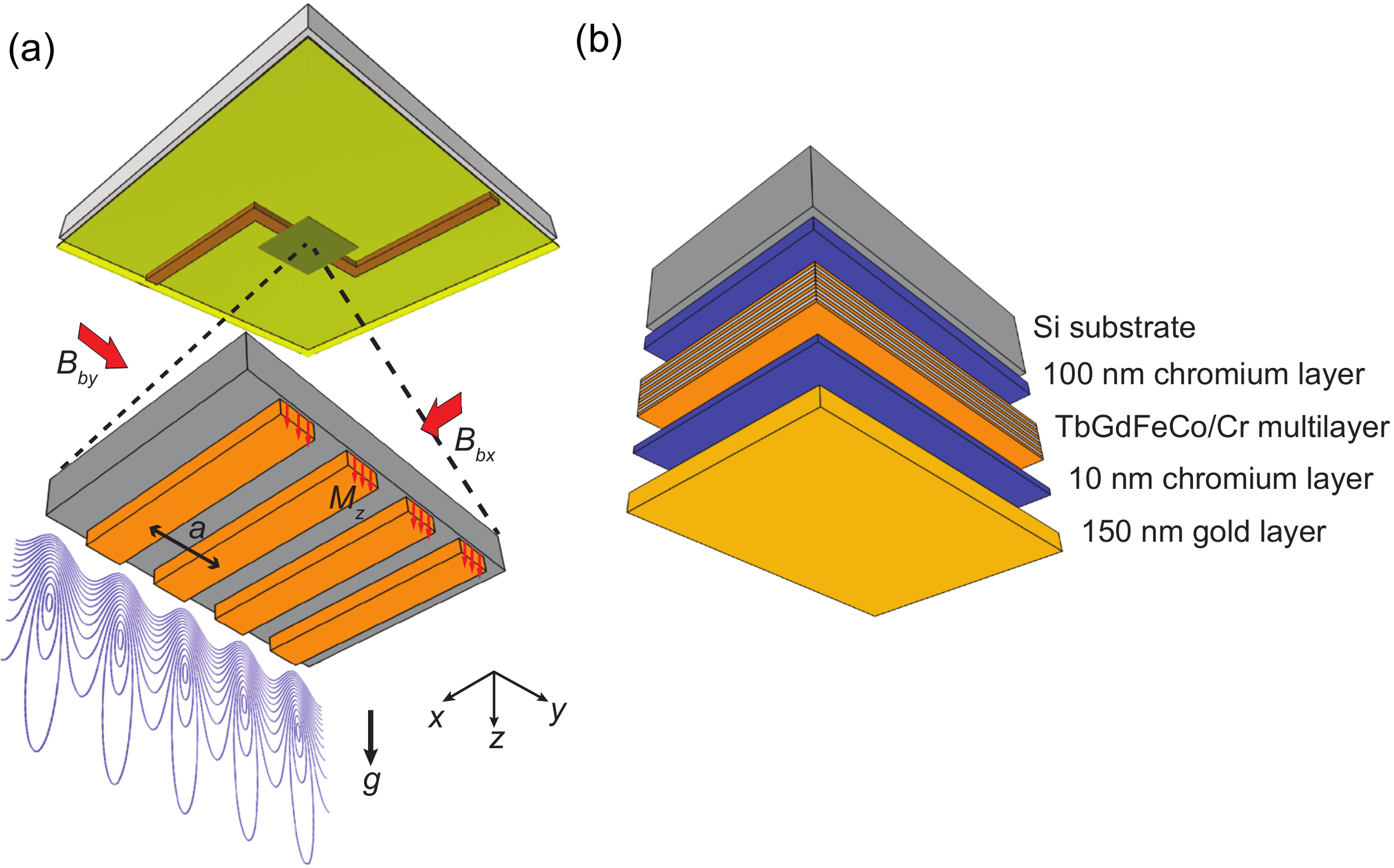} 
		\caption{(Colour online) (a) Schematic of the one-dimensional magnetic lattice created by an array of perpendicularly magnetized parallel magnets with period $a$ and bias fields $B_{bx}, B_{by}$ along the $x$ and $y$ directions. The current-carrying $Z$-wire is shown above the magnetic lattice structure. The contour lines are equipotentials calculated for the parameters: 4$\pi M_z$ = 3 kG, $a = 10$ $\mu$m, $t = 1$ $\mu$m, $\omega_{rad}/2\pi =7.5$ kHz, $\omega_{ax}/2\pi =$ 260 Hz. (b) Multilayered structure of the magnetic film.}
                       \vspace{-1.5em}                
		\label{figure1}
	\end{center}
\end{figure}

For an infinite 1D array of infinitely long, parallel, perpendicularly magnetised magnets with bias fields $B_{bx}$, $B_{by}$ along the $x, y$ directions and no confinement along the (axial) $x$-direction (Fig. \ref{figure1}), the magnetic field components for distances $z >> a/2\pi$ from the surface can be approximated by \cite{Ghanbari06}
\begin{equation}
[B_x;B_y;B_z]\approx[B_{bx};B_0sin(ky)e^{-kz}+B_{by};B_0cos(ky)e^{-kz}]
\end{equation}
where $k = 2\pi/a$, $a$ is the lattice period, $B_0 = 4M_z(e^{kt}-1$) (in Gaussian units), $M_z$ is the magnetization in the $z$ (perpendicular) direction, $t$ is the thickness of the magnets, and $z = 0$ at the top surface of the magnets. The magnetic minima are located at 
\begin{equation}
z_{min}=\frac{a}{2\pi}ln\left(\frac{B_0}{|B_{by}|}\right)
\end{equation}
from the chip surface. The trap bottom given by $B_{min}=|B_{bx}|$ can be adjusted by the bias field $B_{bx}$ to prevent loss of atoms by Majorana spin-flips.

The barrier heights of the traps along the confining $(y,z)$ radial directions are \cite{Ghanbari06}
\begin{align}
&\Delta B_{by}=(B^2_{bx}+4B^2_{by})^{1/2}-|B_{bx}|\nonumber\\
&\Delta B_{bz}=(B^2_{bx}+B_{by}^2)^{1/2}-|B_{bx}|
\end{align}
The trap frequencies for an atom of mass $m$ in a low field-seeking state ($m_F g_F > 0$) confined in a harmonic potential are given by \cite{Ghanbari06}
\begin{equation}
\omega_y=\omega_z=\frac{2\pi}{a}\left(\frac{m_Fg_F\mu_B}{m|B_{bx}|}\right)^{1/2}|B_{by}|
\end{equation}

The above analytic expressions are derived for an infinitely large number of infinitely long magnetic traps with no confinement along the axial ($x$) direction and provide useful scalings for the various magnetic lattice parameters. In practice, our finite magnetic lattice consists of a 1D 10 $\mu$m-period array of one-thousand 10 mm-long traps with weak confinement along the axial direction. To determine the lattice potentials for this finite magnetic lattice we use numerical simulations based on the RADIA code \cite{radia}, as described in Section~IIIA. \\

{\bf{B. Model for RF spectra}}\vspace{0.2cm}

To fit the RF spectra we use a self-consistent Hartree-Fock mean-field model for the BEC plus thermal cloud similar to that used by Gerbier et al. \cite{Gerbier04}, and Whitlock et al. \cite{Whitlock09}. The model accounts for the interaction among atoms in the BEC and in the thermal cloud and the mutual interaction between them, but neglects the kinetic energy of the condensate fraction via the Thomas-Fermi approximation and effects of gravity sag in the tight magnetic traps. The equilibrium condensate density is given by
\begin{equation}
n_c(r)=\mathrm{Max}\{\frac{1}{g}[\mu-V_{ext}(r)-2gn_{th}(r)];0\}
\label{eq.ncr}
\end{equation}
where $n_{th}(r)=\mathrm{Li}_{3/2}[\mathrm {exp}(-|\mu-V_{eff}(r)|)/k_BT]/\lambda^3_{dB}$ is the density distribution of the thermal cloud, $V_{eff}(r)=V_{ext}(r)+2g[n_{th}(r)+n_c(r)]$, $V_{ext}(r)=\frac{1}{2}m\omega^2r^2=\frac{1}{2}m(\omega_x^2x^2+\omega_y^2y^2+\omega_z^2z^2)$ is the confining harmonic potential, and $\mu$ is the chemical potential. Li$_{3/2}[z]$ is the polylogarithmic function with base 3/2, $\lambda_{dB}$ is the thermal de Broglie wavelength, $g=\frac{4\pi\hbar^2a_s}{m}$ is the mean-field coupling constant and $a_s$ is the $s$-wave scattering length. 

In this analysis the atom clouds in the elongated magnetic trap potentials with frequencies $\omega_y = \omega_z = \omega_{rad}$ and $\omega_x= \omega_{ax}$ may be replaced by a spherical trap with a geometric mean trap frequency $\bar{\omega}=(\omega_{rad}^2\omega_{ax})^{1/3}$. To determine $n_c(r)$ eq. \ref{eq.ncr} is solved iteratively for a given temperature $T$, chemical potential $\mu$ and mean-field coupling constant $g$. The resonance condition for RF-induced $\Delta m_F = \pm 1$ spin-flip transitions is $hf'=\mu_B|g_FB|=\frac{1}{|m_F|}\frac{1}{2}m\bar{\omega}^2r'^2$, where $r'^2=(\frac{\omega_x}{\bar{\omega}}x)^2+(\frac{\omega_y}{\bar{\omega}}y)^2+(\frac{\omega_z}{\bar{\omega}}z)^2$, $f'=f-f_0$, $f$ is the applied RF frequency and $f_0$ is the trap bottom. The atomic density distribution as a function of frequency $f'$ is then obtained from the resonance condition and by determining the number of atoms in a spherical shell of volume increment $4\pi r'^2dr'$ that are resonant with an RF knife of frequency $f'$. For a pure BEC, the atomic density distribution as a function of frequency $f'$ is
\begin{align}
n_c(f') &=\mathrm{Max}\{\frac{1}{g}[\mu-V_{ext}(r')]4\pi r'^2(df'/dr')^{-1}; 0\}\nonumber\\
 &=A \mathrm {Max}\{f'^{1/2}(\frac{\mu}{|m_F|h}-f');0\}
\end{align}
where $A=4\sqrt{2}\pi^2\left(\frac{m_F^3}{mh}\right)^{1/2}\frac{1}{\bar{\omega}^3a_s}$. This frequency distribution for a pure BEC has an asymmetric shape originating from the inverted-parabolic (Thomas-Fermi) spatial distribution for a BEC and a base-width $\mu/(|m_F|h)$, which provides a measure of the chemical potential and hence the number of atoms in the condensate.

\begin{figure}[htbp]
	\begin{center}
		\subfigure[\textit{} ]{\label{}\includegraphics[width=.35\textwidth]{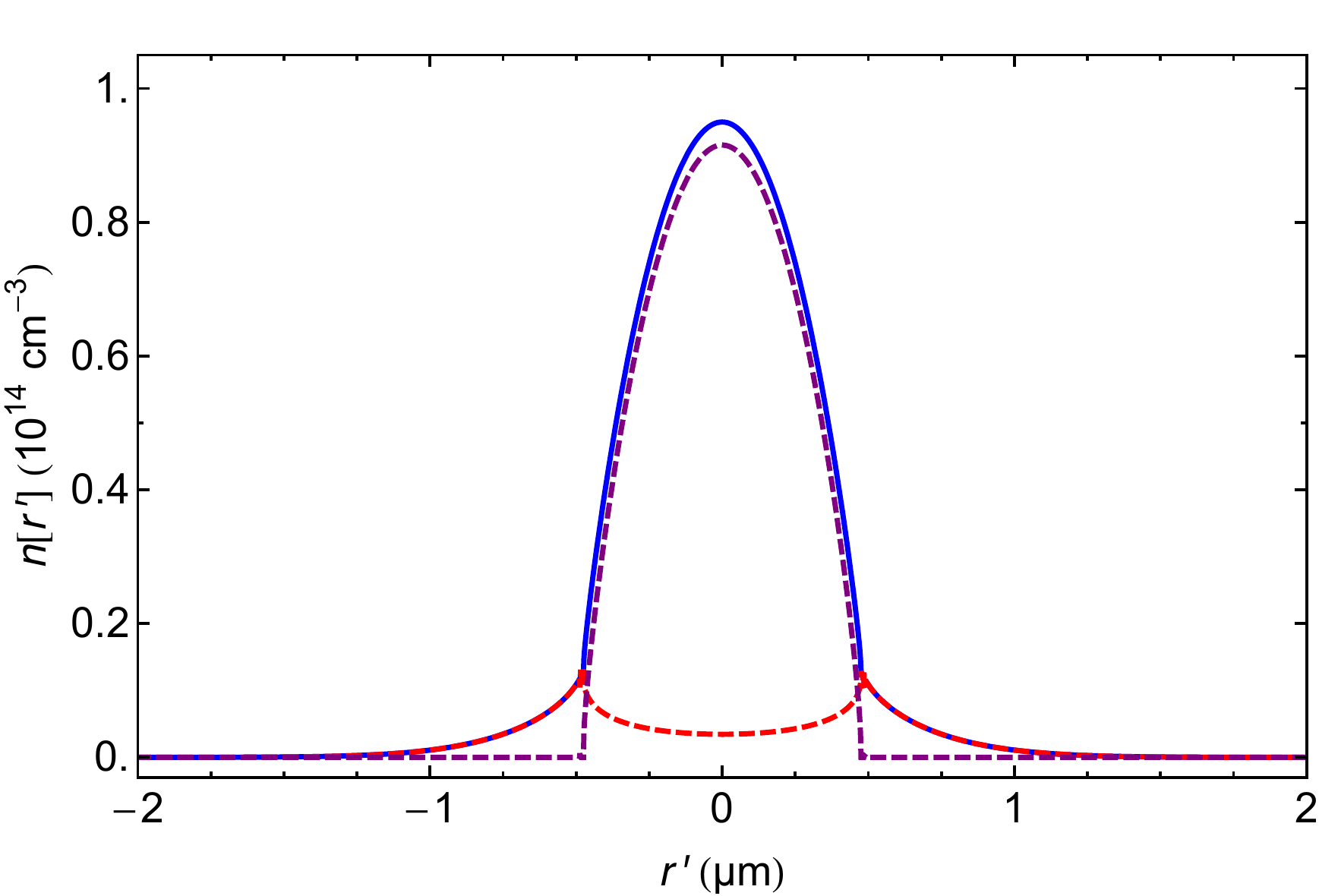}}\vspace{-0.6cm}
                  \subfigure[\textit{} ]{\label{}\includegraphics[width=.35\textwidth]{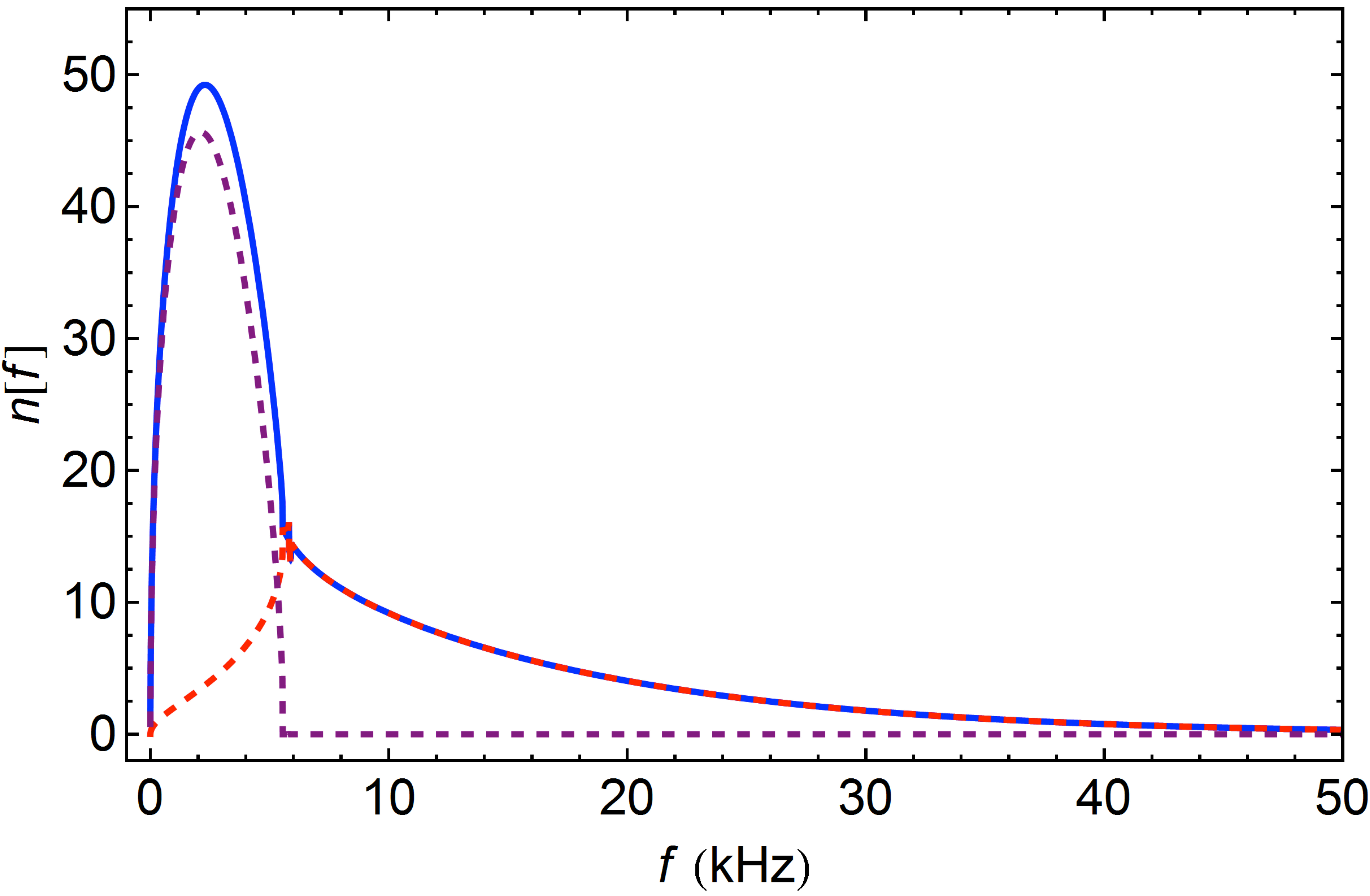}}\vspace{0cm}
		\caption{(Colour online) Calculated (a) atom density versus $r'$ and (b) RF spectrum, showing the contribution of the BEC alone (dashed purple line), the thermal cloud alone (dashed red line) and the sum of the two (blue solid line) for $^{87}$Rb $|F = 1, m_F=-1>$ atoms confined in a magnetic trap. Parameters: $T = 0.5$ $\mu$K, $\mu/h $= 7.5 kHz, $\bar{\omega}/2\pi = 2.40$ kHz, $a_s = 5.3$ nm.}
                       \vspace{-0.9cm}                
		\label{figure2}
	\end{center}
\end{figure}

Figure~\ref{figure2} shows the calculated atom density profile and RF spectrum for a BEC and thermal cloud and the sum of the two for $^{87}$Rb $|F=1,m_F=-1>$ atoms confined in a magnetic trap for the parameters given in the caption. The narrow BEC Thomas-Fermi component is represented by the dashed purple lines and the broad thermal cloud components by the dashed red lines. These distributions show the repulsion of the thermal cloud by the BEC.
\section{III. EXPERIMENTAL}\vspace{-0.4cm}
{\bf {A. Magnetic lattice atom chip}}\vspace{0.2cm}

The magnetic lattice atom chip consists of a one-dimensional perpendicularly magnetized 10 $\mu$m-period grooved TbGdFeCo structure (Fig. \ref{figure1}(a)) mounted on current-carrying wires. Details of the magnetic microstructure and atom chip fabrication have been reported previously \cite{Singh08,SJose13} and only a brief description is given here.

A 300 $\mu$m-thick, 35$\times$35 mm$^2$ silicon wafer is etched with a periodic microstructure consisting of one-thousand 10 mm-long parallel grooves with 10 $\mu$m spacing. The grooved microstructure is coated with six layers of 160 nm-thick magneto-optical Tb$_6$Gd$_{10}$Fe$_{80}$Co$_4$ film, separated by 100 nm-thick non-magnetic chromium layers (Fig. \ref{figure1}(b)), using a magnetron sputtering system operated at a base pressure of typically 10$^{-8}$ Torr. The grooves are sufficiently deep ($>$20 $\mu$m) compared with the lattice period that the magnetic film at the bottom of the grooves has almost no influence on the magnetic potentials. A 150 nm-thick gold film is deposited on top of the multi-layer magnetic structure which gives good optical reflectivity ($> $95 \% at 780 nm) for the mirror MOT and for reflective absorption imaging of the atom clouds. The magnetic microstructure is aligned with and glued on top of the wire pattern consisting of two U-wires and a Z-wire (both 5 mm $\times$ 1 mm $\times$ 0.5 mm) \cite{Singh08}.

Hysteresis loops obtained using a SQUID magnetometer show that the multilayer Tb$_6$Gd$_{10}$Fe$_{80}$Co$_4$ film has a remanent magnetization 4$\pi M_z$ = 3 kG and a coercivity $H_c$ = 6 kOe. Scanning electron microscope measurements indicate a grain size of about 40 nm \cite{Wang05}. The surface roughness of the magnetic microstructure measured by an atomic force microscope is less than 20 nm. The TbGdFeCo microstructure is magnetized in a direction perpendicular to the surface of the film and analysed using a magnetic force microscope (MFM) which measures the second derivative of the magnetic field along the $z$ (vertical) direction \cite{hughes97}. The MFM images indicate that the microstructure produces a periodic magnetic field with a period of 10 $\mu$m.

\begin{figure}[htbp]
	\begin{center}
		\subfigure[\textit{} ]{\label{}\includegraphics[width=.25\textwidth]{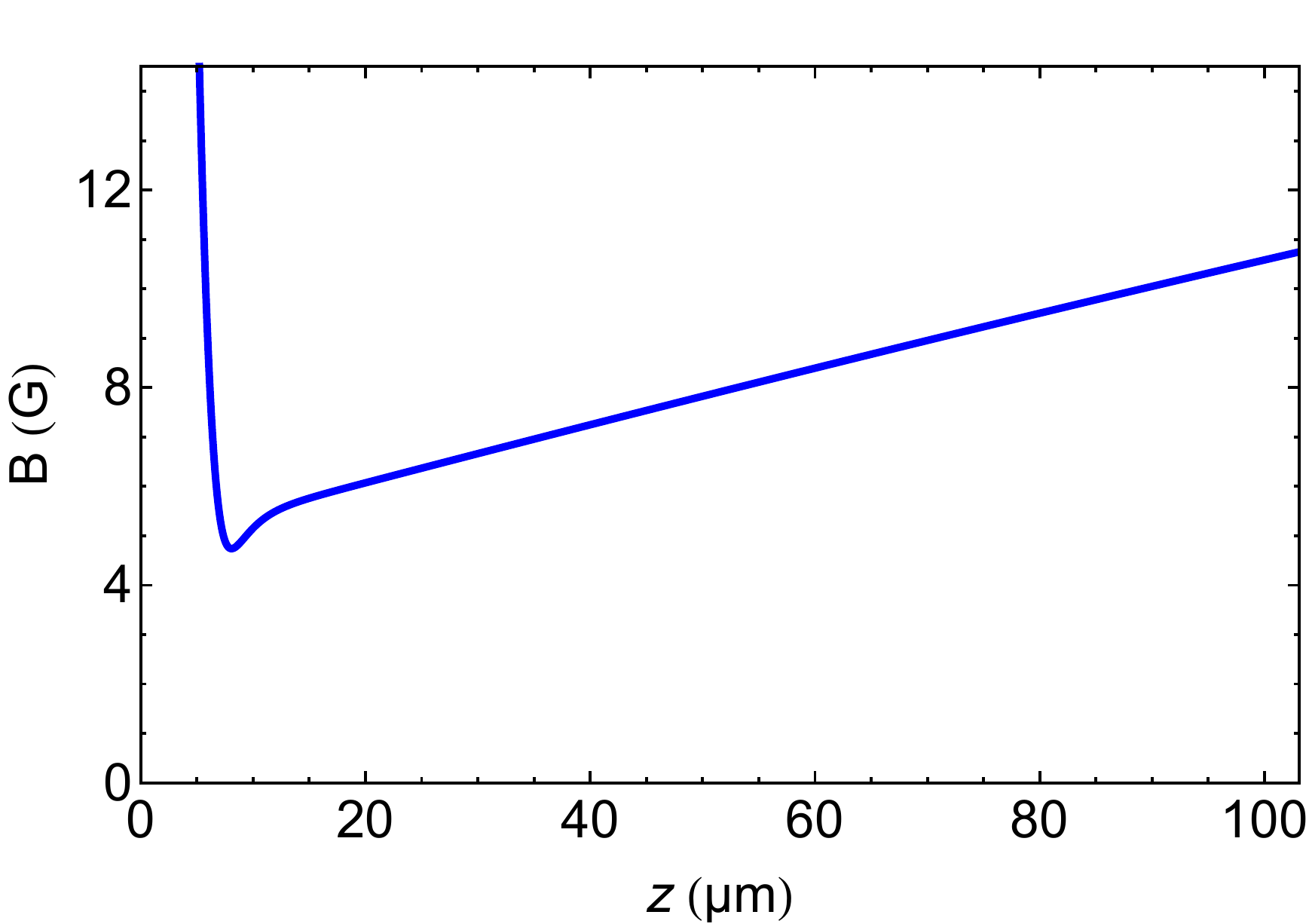}}\vspace{-0.6cm}
                  \subfigure[\textit{} ]{\label{}\includegraphics[width=.25\textwidth]{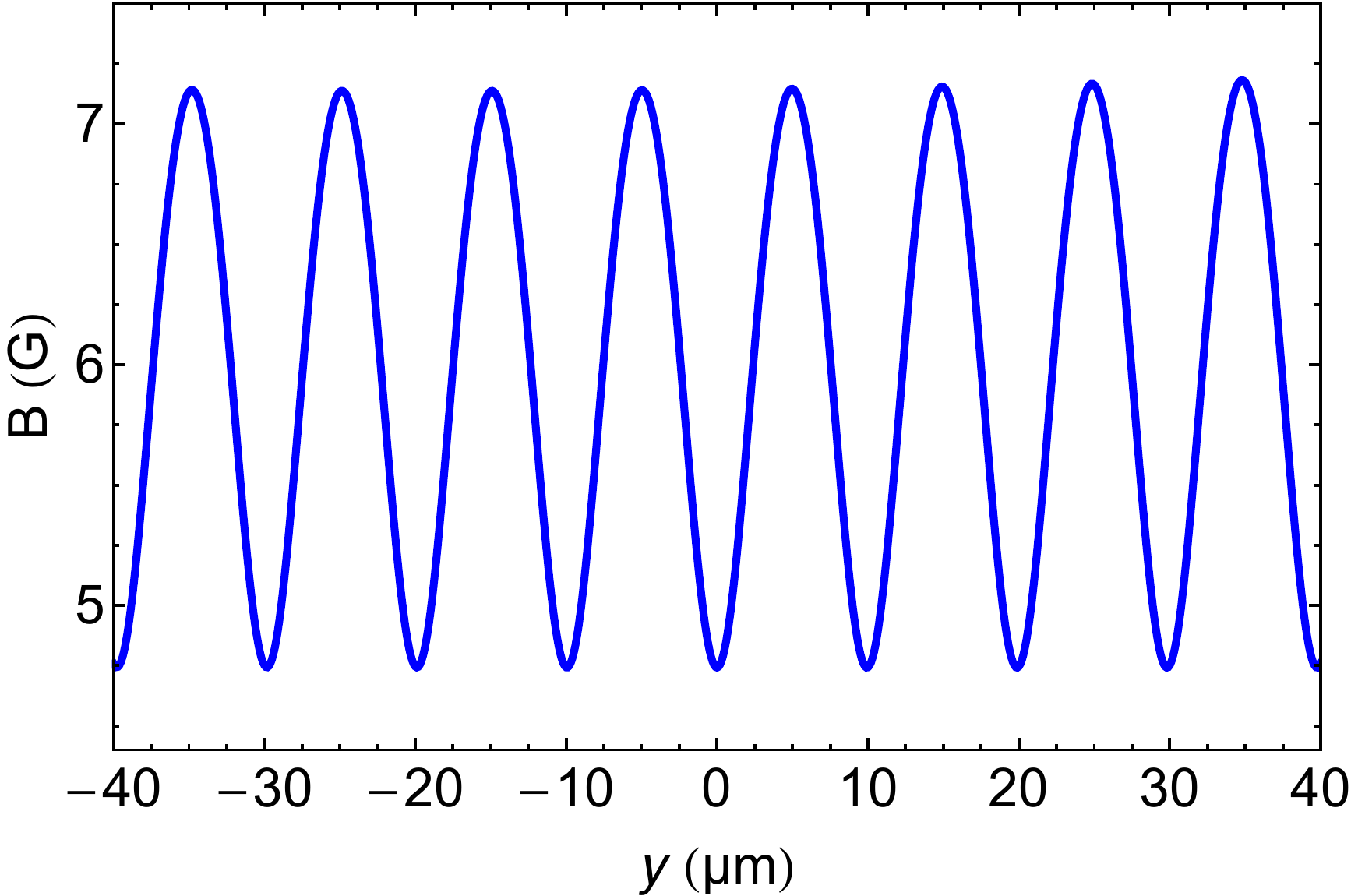}}\vspace{-0.6cm}
                 \subfigure[\textit{} ]{\label{}\includegraphics[width=.25\textwidth]{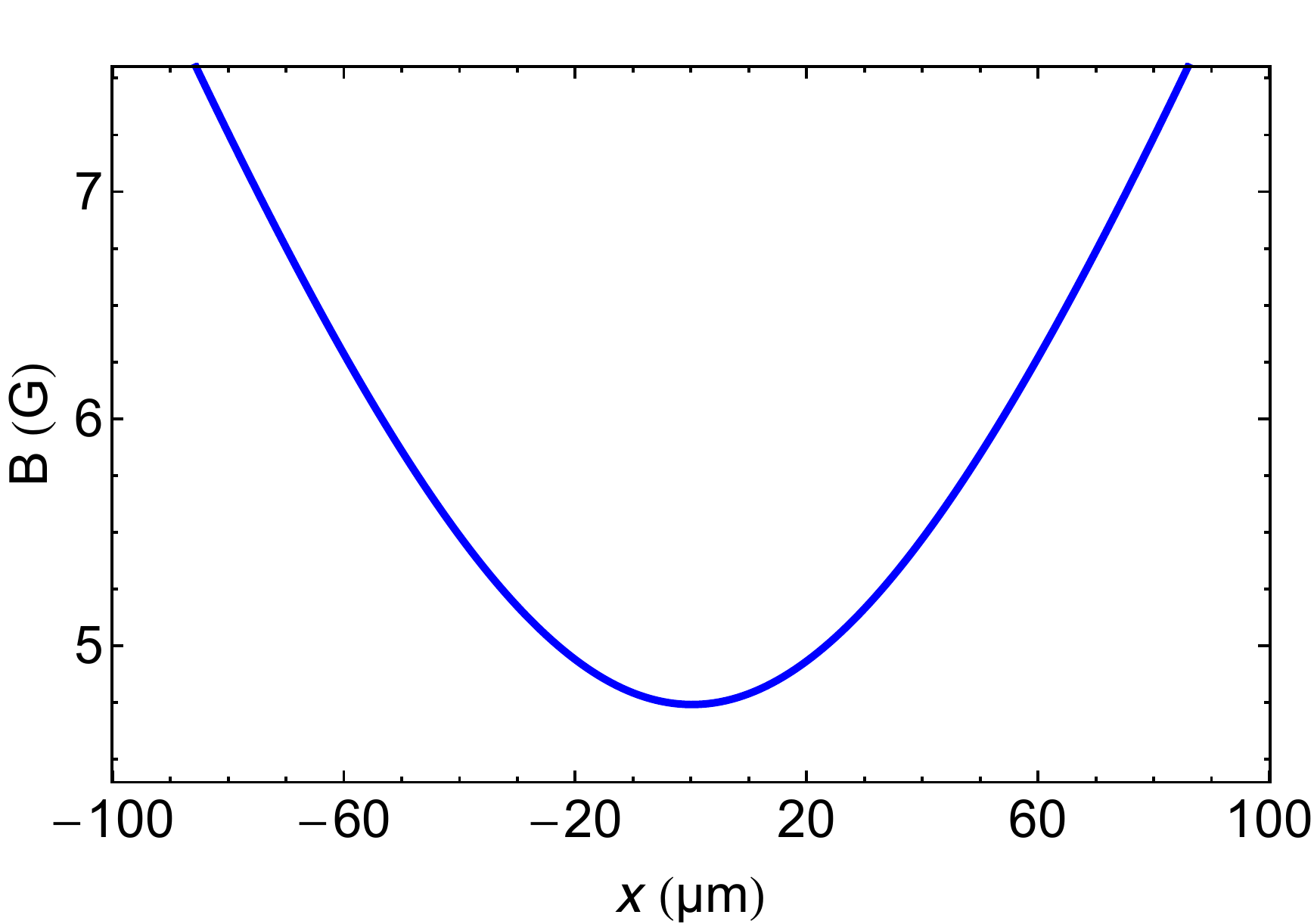}}\vspace{0cm}
		\caption{Simulations of the magnitude of the magnetic field along the (a) $z$ (b) $y$ and (c) $x$ directions of the lattice microtraps using the RADIA code [19]. Parameters: 4$\pi M_z = 3$ kG, $t = 0.96$ $\mu$m, $a =10$ $\mu$m, $I_z =17 $ A, $B_{bx}=51$ G, $B_{by}^c =0$.}
                       \vspace{-2.5em}                
		\label{figure3}
	\end{center}
\end{figure}

The $y$-bias field required to create the magnetic lattice potential is provided by the end-wires of the current-carrying $Z$-wire along the $x$-direction plus an additional magnetic
field $B_{by}^c$ provided by external coils, while the axial confinement for the traps is provided by the central region of the $Z$-wire along the $y$-direction \cite{SJose13}. Figure \ref{figure3} shows simulations of the magnitude of the magnetic field along the $z, y$ and $x$ directions of the magnetic lattice traps for the parameters given in the caption. The magnetic field in the $z$-direction consists of a narrow trapping region together with a broad trapping region that falls off slowly with distance $z$ (Fig. \ref{figure3}(a)) which is advantageous for trapping a large volume of atoms,. The barrier heights of the lattice traps are $\Delta B_{by}$ = 2.4 G, or 82 $\mu$K for $|F = 1, m_F =-1 >$ atoms, and $\Delta B_{bz} >$ 5 G, or $>$ 180 $\mu$K. The potential minima are located at $z_{min} = 8$ $\mu$m from the chip surface.
\vspace{0.2cm}

{\bf {B. Trap frequencies}}\vspace{0.2cm}

\begin{figure}[htbp]
	\begin{center}
		\subfigure[\textit{} ]{\label{}\includegraphics[width=.3\textwidth]{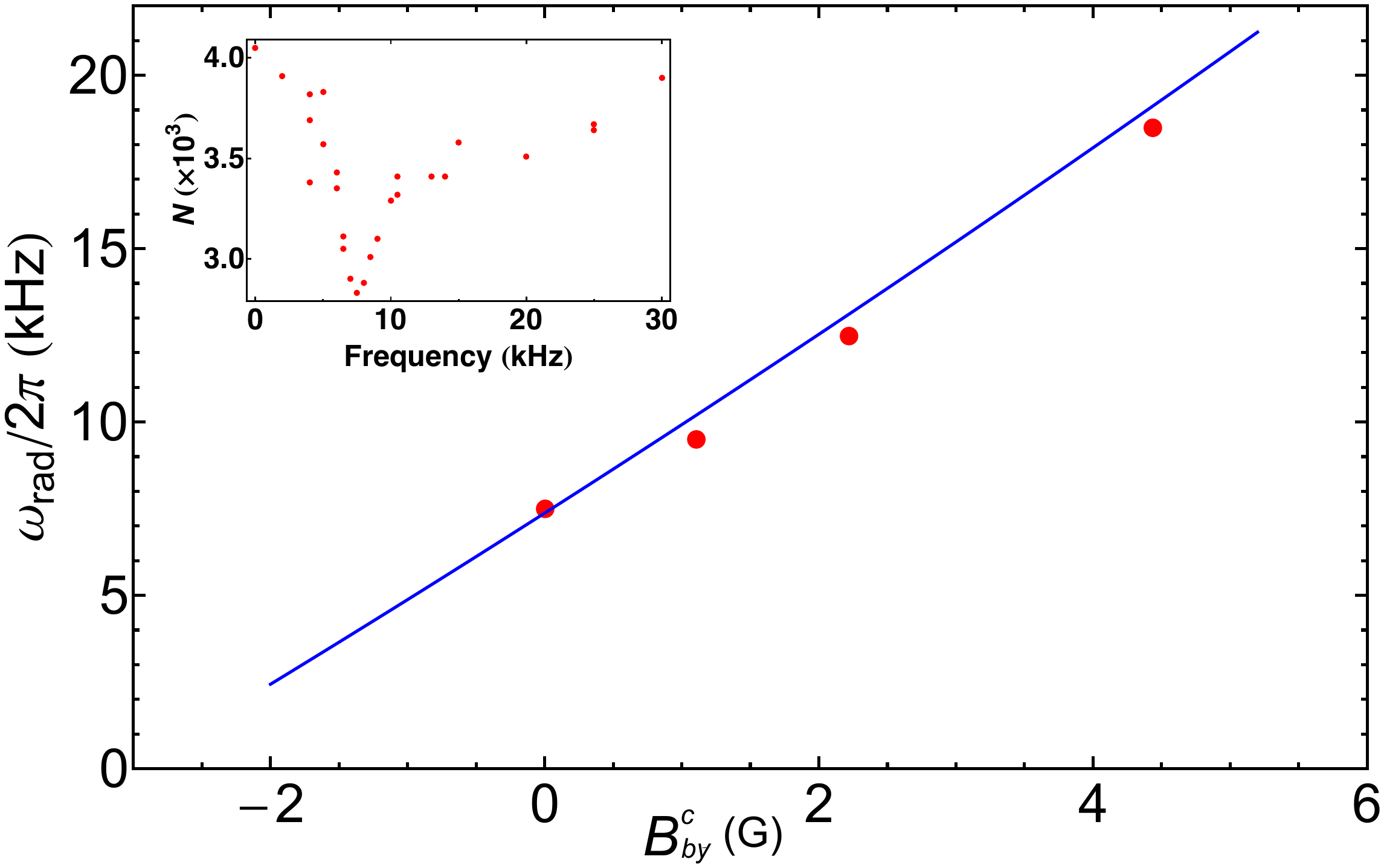}}\vspace{-0.6cm}
                  \subfigure[\textit{} ]{\label{}\includegraphics[width=.305\textwidth]{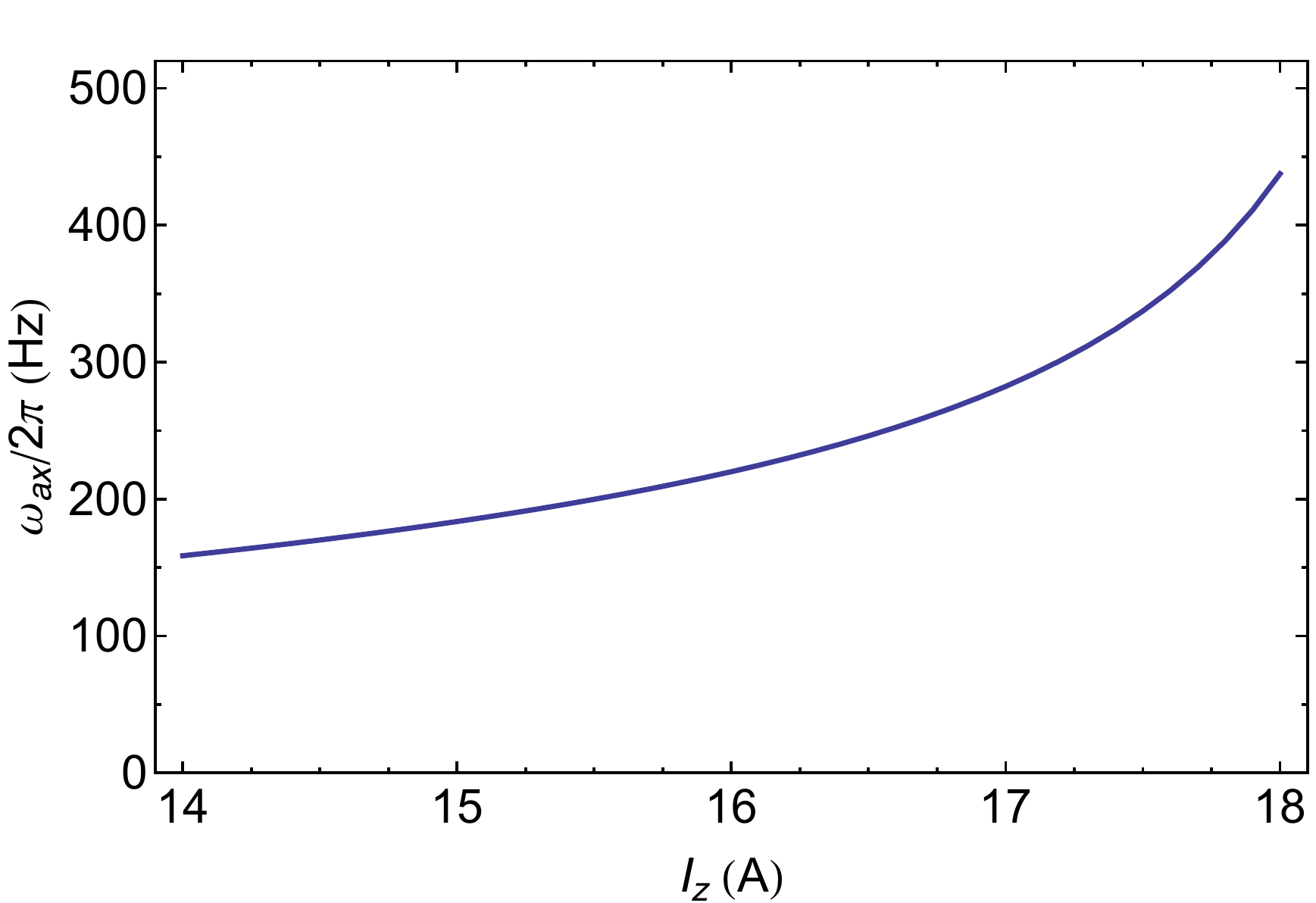}}\hspace{0.2cm}\vspace{-0.2cm}
		\caption{(a) Radial trap frequency of the magnetic lattice traps versus $B_{by}^c$ provided by external coils for $I_z = 17$ A, $B_{bx} = 51$ G measured by parametric heating. The solid line represents the calculated radial trap frequency over the range used in the experiments. The inset shows the parametric heating resonance for $B_{by}^c=0$. (b) Calculated axial trap frequency versus $I_z$ for $B_{by}^c=0$, $B_{bx}=51$ G.}
                       \vspace{-2.5em}               
		\label{figure4}
	\end{center}
\end{figure}

To determine the radial trap frequencies in the tight magnetic lattice traps, we use the $in$-$situ$ technique of parametric heating \cite{Jauregui01}. A 250 ms-long oscillating pulse is applied through one of the $U$-wires to modulate the stiffness constant of the traps. When the applied frequency matches the trap frequency the atoms absorb energy from the external source resulting in a rise in temperature and a loss of trapped atoms (Fig. \ref{figure4}, inset). For a $Z$-wire current $I_z$ = 17 A and $B_{bx} $= 51 G, the measured radial trap frequency $\omega_{rad}/2\pi$ = 7.5 $\pm$ 0.1 kHz.

The radial trap frequency can be varied while keeping the axial trap frequency constant by applying a $y$-bias field, $B_{by}^c$, created by external coils (Fig. \ref{figure4}(a)). The measured radial trap frequencies agree with the simulated values (solid line in Fig. \ref{figure4}(a)) with a mean rms difference of 2\% over the range investigated.

Figure \ref{figure4}(b) shows a plot of the calculated axial trap frequency versus $I_z$ for $B_{by}^c$= 0 and $B_{bx}$ = 51 G. For $I_z$ = 17 A, the calculated $\omega_{ax}/2\pi$ = 260 Hz.
\vspace{0.2cm}

{\bf {C. Atom cooling and loading magnetic lattice}}\vspace{0.2cm}

The atom cooling and trapping cycle begins with $^{87}$Rb atoms released from a dispenser into a standard four-beam mirror magneto-optical trap (MMOT) in which two of the beams are retro-reflected at 45$^o$ to the gold surface on the chip \cite{Singh08}. The beams consist of combined trapping and re-pumper beams in which the trapping laser is detuned 14 MHz below the $F = 2 \rightarrow F' = 3$ cycling transition and the re-pumper laser is locked to the $F=1 \rightarrow F'=2$ transition. We trap typically $2\times 10^8$ atoms in 28 s in the MMOT at $\sim$1.2 mm below the chip surface. The atoms are then transferred to a compressed MOT by passing 20 A through a $U$-wire and applying a bias field $B_{bx}$ = 12 G to create a quadrupole magnetic trap. This is followed by a polarization gradient cooling stage resulting in $\sim 1\times 10^8$ atoms cooled to 30-40 $\mu$K.

Next, the atoms are optically pumped into the $|F = 1, m_F = -1 >$ low field-seeking state which is chosen for trapping in the magnetic lattice because of its three-times smaller three-body loss rate \cite{Burt97,Soeding99} and weaker magnetic confinement compared with the $|F = 2, m_F = 2 >$ state. The optical pumping is performed by applying 4 ms-long $\sigma ^-$-polarized pulses resonant with the $F = 2 \rightarrow F' = 2$ and $F = 1 \rightarrow F' = 2$ transitions to first pump atoms into the $|F=2,m_F =-2>$ dark state and then switching off the $F=1\rightarrow F'=2$ pulse to allow the atoms to accumulate in the $|F= 1, m_F = -1 >$ state. The $\sigma^{-}$-pulse contains a small $\pi$-polarized component to remove atoms from the $|F = 2, m_F = -2 >$ dark state.

The $|F = 1, m_F = -1 >$ atoms are then transferred to a $Z$-wire Ioffe-Pritchard magnetic trap formed by passing a current of $I_z$ = 35 A, increasing the $x$-bias field to $B_{bx}$ = 10 G and
applying a $y$-bias field of $B_{by}^c$= 7 G from external coils. The cloud is then compressed by ramping $I_z$ and $B_{bx}$ up to 37 A and 51 G and raising $B_{by}^c$ to 9.5 G in 100 ms, resulting in $\sim 5 \times 10^7$ atoms trapped in the $Z$-wire trap at $\sim$ 600 $\mu$m below the chip surface with a trap lifetime of about 25 s. RF evaporative cooling is then performed for 10 s in the $Z$-wire trap by ramping the frequency of the RF field down to a final evaporation frequency of $\sim$ 3 MHz leaving $\sim 3 \times 10^6$ atoms in the $Z$-wire trap at 10-15 $\mu$K.
$I_z$ is ramped from 37 A down to 17 A in 100 ms keeping $B_{bx}= 51$ G with $B_{by}^c= 0$, where the $Z$-wire trap merges smoothly with the magnetic lattice traps allowing $\sim 3 \times 10^5$ $|F = 1,m_F = -1 >$ atoms to be loaded into about 100 lattice sites located 8 $\mu$m from the chip surface. A second evaporation ramp is then carried out from 7 MHz down to a final frequency $f_f\sim$ 5 MHz in 1.5 s to further cool the atoms in the magnetic lattice.\vspace{0.2 cm}

{\bf {D. Absorption imaging and site-resolved radio frequency spectroscopy}}\vspace{0.2cm}

\begin{figure}[htbp]
	\begin{center}
            \subfigure[\textit{} ]{\label{}\includegraphics[width=.47\textwidth]{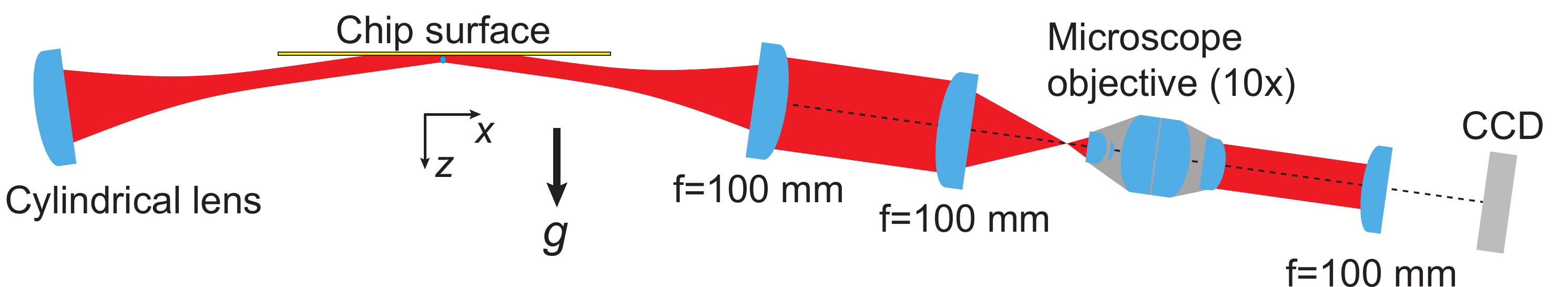}}\vspace{-0.2cm}
            \subfigure[\textit{} ]{\label{}\includegraphics[width=.47\textwidth]{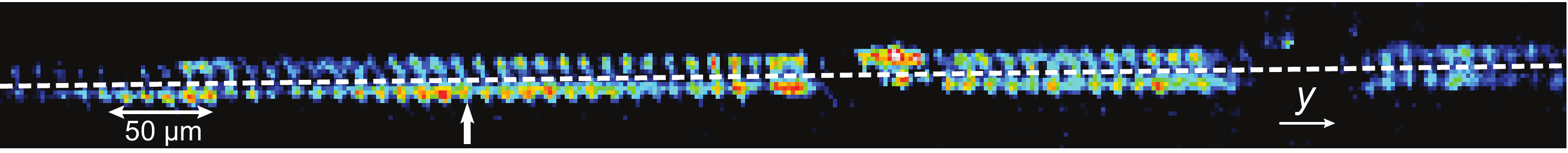}}\vspace{0cm}		
\caption{(Colour online) (a) Schematic of the reflection absorption imaging system. (b) Absorption image for an array of about 100 clouds of $^{87}$Rb $|F = 1, m_F = -1 >$ atoms trapped in the 1D 10 $\mu$m-period magnetic lattice after evaporative cooling to a trap depth $\delta f = (f_f-f_0) = 100$ kHz. Images are
produced both prior to and after reflection of the imaging beam from the chip surface (indicated by the horizontal dashed line). The gaps in the absorption signals for some of the sites correspond to regions where the reflectivity of the gold mirror is low due to contamination by rubidium atoms. The vertical white arrow indicates the lattice site (site 38) at which the RF spectra in Figs. \ref{figure6}, \ref{figure9}, \ref{figure12} were recorded. The effective pixel size is 2.0 $\mu$m.}
                       \vspace{-1.5em}                
		\label{figure5}
	\end{center}
\end{figure}

For absorption imaging, the atoms are pumped into the $|F = 2, m_F=+2>$ state and a $\sigma^+$-polarized imaging beam tuned to the $F=2\rightarrow F=3$ cycling transition is focused by a 700 mm-focal length plano-convex cylindrical lens into a light sheet \cite{Armijo10} to produce a uniform imaging beam across all occupied sites of the magnetic lattice (Fig. \ref{figure5}(a)). The light transmitted by the atoms is imaged by a 100 mm-focal length, 50.8 mm-diameter achromatic lens positioned next to the vacuum viewport. To minimize aberration and vignetting a second identical lens is positioned 200 mm away (in a 4f configuration). The image is magnified using a commercial microscope objective (10$\times$ aplanat) and a 100 mm-focal length, 50.8 mm-diameter tube lens to create an image on the CCD camera. With the 100 mm-focal length tube lens the magnification is 6.5 and the effective pixel size 2.0 $\mu$m. The 100 mm working distance and the 45 mm-diameter effective aperture give a numerical aperture $NA$ = 0.22 and a theoretical diffraction-limited resolution (Rayleigh criterion) $R = 0.61\lambda/NA$ = 2.2 $\mu$m. The actual resolution measured from the width of the BEC images of individual lattice sites is 4 $\mu$m.

The images are recorded using reflection absorption imaging \cite{Armijo10} along the $x$-direction parallel to the long axis of the elongated atom clouds. The imaging beam is sent at a slight angle ($\sim 2^o$) to the reflecting gold surface of the chip, resulting in images both prior to and after reflection of the imaging beam at the CCD camera, which is operated in frame-transfer mode. The deep ($>$20 $\mu$m) grooves along the direction of the imaging beam affect the reflection of the imaging beam from the gold surface in the region of the atom clouds. A small misalignment of the imaging beam from the $x$-direction allows sufficient reflection to obtain absorption signals from the elongated atom clouds.  

Figure \ref{figure5} (b) shows a reflection absorption image for a periodic array of clouds of $^{87}$Rb atoms trapped in about 100 sites of the 1D 10 $\mu$m-period magnetic lattice after evaporative cooling to a trap depth $\delta f = (f_f - f_0)$ = 100 kHz (where $f_f$ is the final evaporation frequency). The atom clouds are resolved in their individual lattice sites, which allows us to perform site-resolved RF spectroscopy measurements. The gaps in the absorption signals for some of the sites correspond to regions where the reflectivity of the gold mirror is low due to contamination by rubidium atoms adsorbed on the gold surface. The site-to-site variations in the absorption signals are also mainly due to variations in the reflectivity of the gold mirror.\vspace{-0.05cm}

The measured positions of the atom clouds indicate that the period of the array is constant to better than 1\%. The separation of the top and bottom absorption images provides a measure of the distance of the atoms from the chip surface, which is 8 $\mu$m, in agreement with the calculated value.

The RF measurements were performed by applying an RF pulse of duration 40 ms after evaporative cooling in the magnetic lattice. The amplitude of the RF pulse was reduced ten times after RF evaporation to minimize power broadening.\vspace{0.2cm}

{\bf {E. Magnetic noise broadening of RF spectra}}\vspace{0.2cm}

The contribution of magnetic noise broadening, mainly at the 50 Hz mains frequency, to the RF spectra is determined by comparing the experimental RF spectra taken over a range of radial trap frequencies (Fig. \ref{figure9}) with theoretical RF spectra generated by convoluting frequency distributions obtained from the self-consistent mean-field model with Gaussian broadening functions of varying width. The theoretical fits are constrained so that the atom number derived from the fit matches the (scaled) atom number determined from absorption imaging measurements (Fig. \ref{figure10}(d)). The scaling factor for the atom number is introduced to allow for the effect of laser light that is scattered from the chip surface without passing through the atom cloud and for imperfections in the absorption imaging process. By comparing the atom number determined from the absorption measurements with the (unconstrained) atom number derived from theoretical fits to the RF spectra taken at large radial trap frequencies where the effect of magnetic noise broadening is relatively small we obtain a scaling factor of 2.8. Comparing the experimental RF spectra with theoretical spectra generated in this way, we obtain a FWHM = 4.3 $\pm$ 0.2 kHz for the Gaussian magnetic noise function. This is to be compared with a peak-to-peak magnetic noise of $\sim$ 5.2 mG, or $\sim$ 3.6 kHz for rubidium $|F=1, m_F=-1>$ atoms, measured on the outside of the UHV chamber using a fluxgate magnetometer with all power supplies and electronics operating.\vspace{-0.6cm}
\section{IV. RESULTS}\vspace{-0.4cm}
{\bf {A. RF spectra for varying trap depth}}\vspace{0.2cm}
\begin{figure}[h]
	\begin{center}
		\includegraphics[width=0.32\textwidth]{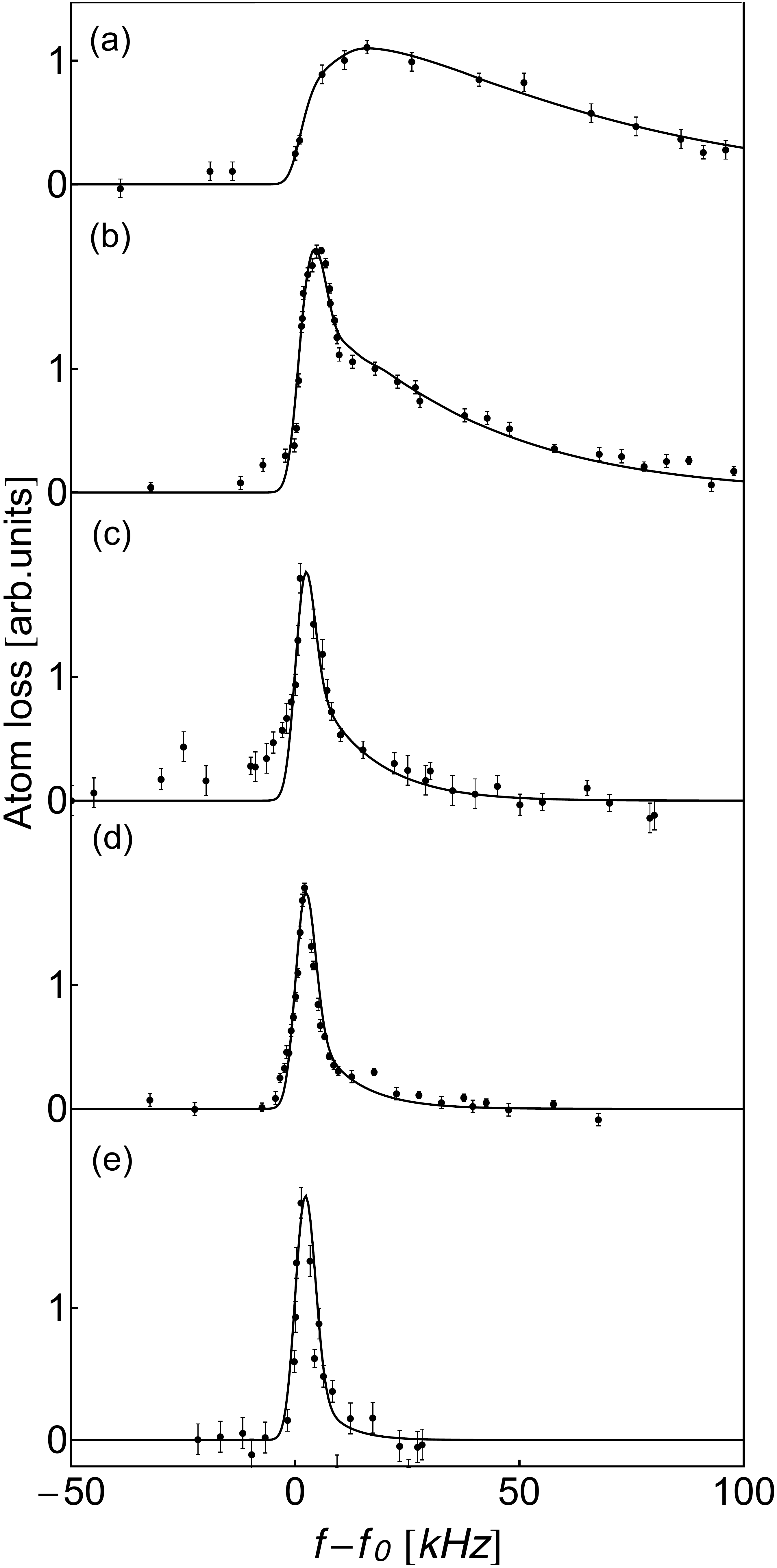} 
		\caption{ RF spectra of loss of atoms in lattice site 38 after evaporative cooling to trap depths $\delta f$ of (a) 600 kHz (b) 400 kHz (c) 200 kHz (d) 100 kHz and (e) 50 kHz, with $\omega_{rad}/2\pi = 7.5$ kHz, $\omega_{ax}/2\pi = 260$ Hz. The points represent mostly single shots. The error bars were estimated from the average standard deviation of the mean for those points for which multiple shots were taken divided by the square root of the number of shots for each point. The solid lines are fits to the data points based on the self-consistent mean-field model for a BEC plus thermal cloud convoluted with a Gaussian magnetic noise function with FWHM = 4.3 kHz as described in the text. The temperatures and atom numbers are (a) $T=2$ $\mu$K, $N_\mu=5350$;     (b)$T=1.3$ $\mu$K, $N_\mu =3430$; (c) $T=0.5$ $\mu$K, $N_\mu =860$; (d) $T=0.38$ $\mu$K, $N_{abs} = 200$; and (e) $T = 0.25$ $\mu$K, $N_{abs} = 160$. The $N_{abs}$ are determined from absorption measurements and the $N_\mu$ from fits to the RF spectra.}
                       \vspace{-1.5em}                
		\label{figure6}
	\end{center}
\end{figure}

\begin{figure}[htbp]
\centering  
 \includegraphics[width=0.32\textwidth]{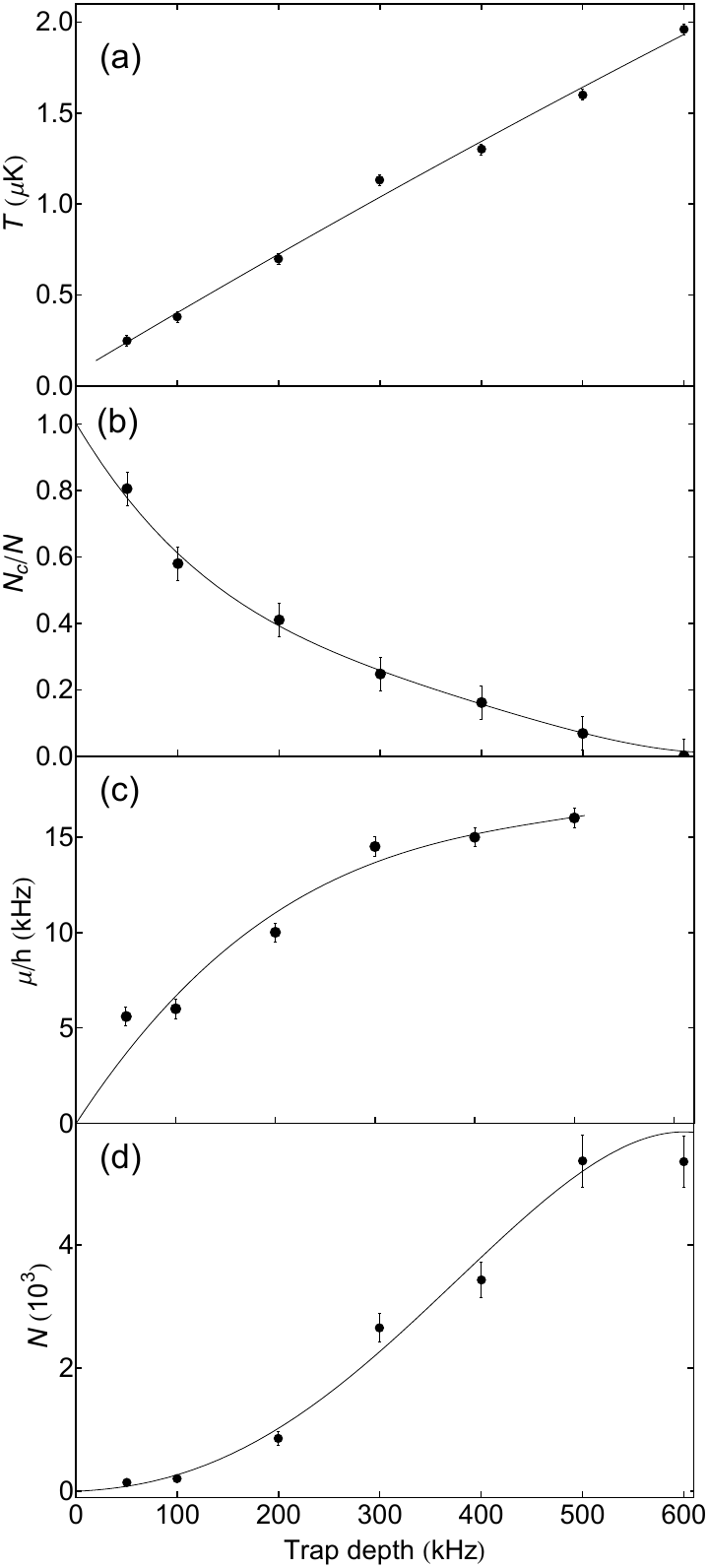} 
\caption[]{Variation of (a) temperature $T$ (b) condensate fraction $N_c/N$ (c) chemical potential $\mu/h$ and (d) atom number per site $N$, with trap depth for lattice site 38 with $\omega_{rad}/2\pi = 7.5$ kHz, $\omega_{ax}/2\pi= 260$ Hz. The points were determined from fits to the RF spectra in Fig. \ref{figure6}, as described in the text. The solid lines are polynomial fits constrained to pass through (b)$N_c/N = 1.0$ (c) $\mu/h = 0$ and (d) $N = 0$ at zero trap depth.}
\label{figure7}
  \vspace{-1.0em}   
\end{figure}

Figure \ref{figure6} shows RF spectra recorded for lattice site 38 after the atoms in the magnetic lattice are evaporatively cooled to trap depths ranging from $\delta f = 600$ kHz down to $\delta f =
50$ kHz, with $\omega_{rad}/2\pi$ = 7.5 kHz, $\omega_{ax}/2\pi$ = 260 Hz. For trap depths $\delta f < $ 600 kHz the RF spectra exhibit bimodal distributions consisting of a narrow BEC component plus a broad thermal cloud component, analogous to those obtained for the density distribution of a BEC plus thermal cloud in conventional time-of-flight measurements. The solid lines in Fig. \ref{figure6} are fits to the data points obtained from the self-consistent Hartree-Fock mean-field model for a BEC plus thermal cloud convoluted with a Gaussian magnetic noise function with FWHM = 4.3 kHz. For the $\delta f = 50$ kHz and 100 kHz spectra, the fits are obtained by fitting the temperature and constraining the atom number so that it matches the scaled atom number determined from the absorption imaging measurements. The primary effect of temperature on the RF spectra is to change both the condensate fraction and the width of the thermal cloud component. For the spectra taken at the larger trap depths ($\delta f > 100$ kHz), the atom number is up to 30 times higher (Fig. \ref{figure7}(d)), and the absorption imaging results are no longer reliable owing to saturation of the absorption signal by the presence of a non-absorbed background light component. Therefore, the spectra for $\delta f >$ 100 kHz are fitted by adjusting both the temperature and the chemical potential, from which the atom number $N_{\mu}$ is determined.\vspace{-0.55cm}
\vspace{0.2cm}

Figures \ref{figure7}(a) and (b) show plots of the temperature and condensate fraction determined from the fits to the RF spectra in Fig. \ref{figure6} as a function of trap depth. With decreasing trap depth, the temperature continues to decrease approximately linearly, down to 0.25 $\mu$K at the lowest trap depth ($\delta f$ = 50 kHz) for which the ideal-gas critical temperature for quantum degeneracy is $T_c ^0$ = 0.59 $\mu$K for $N = 160$ atoms per site. The decrease in temperature is accompanied by an increase in condensate fraction, which for $\delta f = 50$ kHz is close (81\%) to that of a pure condensate. The chemical potential and atom number per site (Figs. \ref{figure7}(c) and (d)) decrease monotonically with trap depth, approaching zero at small trap depths.

\vspace{0.2cm}
{\bf {B. RF spectra for various sites across the magnetic lattice}}\vspace{0.2cm}

Our light-sheet imaging scheme allows site-resolved RF spectra to be recorded simultaneously for nearly all occupied sites across the magnetic lattice, with a total acquisition time of around one hour. Since the absorption signals show quite large site-to-site variations across the lattice, due mainly to variations of reflectivity of the gold mirror (Fig. \ref{figure5}), this set of RF spectra were analysed by fitting both temperature and chemical potential, from which the atom number per site $N_\mu$ was determined.
\begin{figure}[h]
	\begin{center}
                  \includegraphics[width=0.35\textwidth]{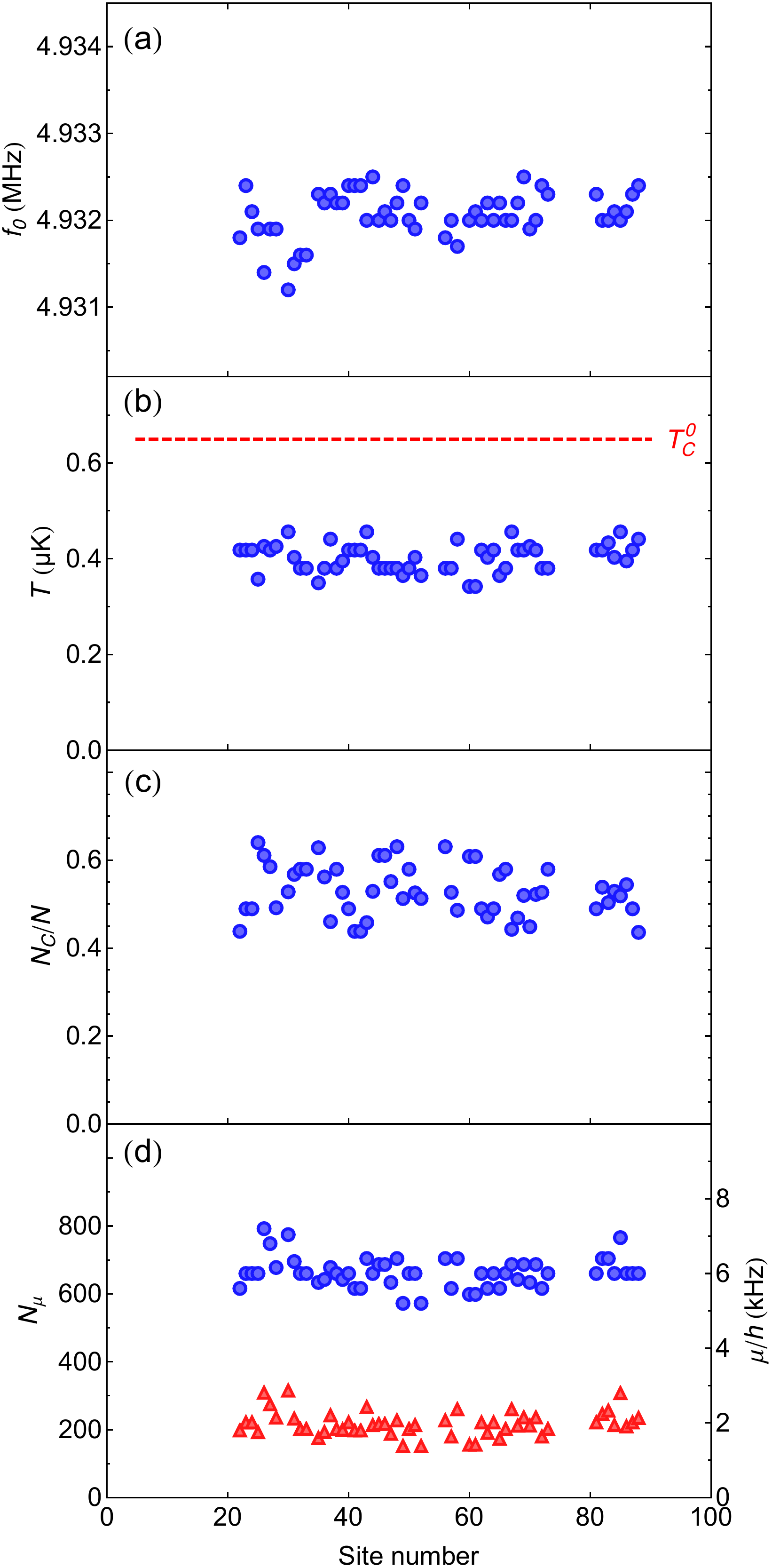} 
		\caption{ (Colour online) (a) Trap bottom $f_0$ (b) atom temperature $T$ (c) condensate fraction $N_c/N$ and (d) chemical potential $\mu/h$ (blue points) and atom number $N_\mu$ (red triangular points) determined from fits to the RF spectra for 54 sites across the central region of the magnetic lattice, with $\delta f = 100$ kHz, $\omega_{rad}/ 2\pi = 7.5$ kHz, $\omega_{ax}/2\pi =260$ Hz. The red dashed line in (b) represents the ideal-gas critical temperature $T_c^0$ for 220 atoms. The mean values and standard deviations of the quantities for the 54 sites are $<$$f_0$$>$ = 4.9323 $\pm$ 0.0003 MHz, $<$T$>$ =  0.40 $\pm$ 0.03 $\mu$K, $<$$N_c/N$$>$ = 0.54 $\pm$ 0.06, $<$$\mu/h$$>$ = 6.0 $\pm$ 0.4 kHz and $<$$N_\mu$$>$ = 220 $\pm$   40. }
                       \vspace{-0.8cm}                
		\label{figure8}
	\end{center}
\end{figure}
\vspace{0.05cm}

Figure \ref{figure8} shows plots of the trap bottom, temperature, condensate fraction, chemical potential and atom number determined from fits to the RF spectra for 54 sites across the central region of the magnetic lattice for a trap depth $\delta f = 100$ kHz. Large condensate fractions (0.54 $\pm$ 0.06) are found for all 54 lattice sites (Fig. \ref{figure8}(c)). The site-to-site variation in the various quantities, given by the standard deviations in the caption to Fig. \ref{figure8}, indicate that the sites are remarkably uniform across the magnetic lattice. In particular, the trap bottoms, which could be accurately determined from the intercepts of the fitted RF spectra with the ($f - f_0$) axis, show site-to-site variations of only $\pm$ 0.3 kHz, or $\pm$ 0.4 mG for $|F = 1, m_F = -1 > $ atoms.\vspace{0.2cm}

{\bf {C. RF spectra for varying trap frequencies}}\vspace{0.2cm}
\begin{figure}[h]
	\begin{center}
		\includegraphics[width=0.30\textwidth]{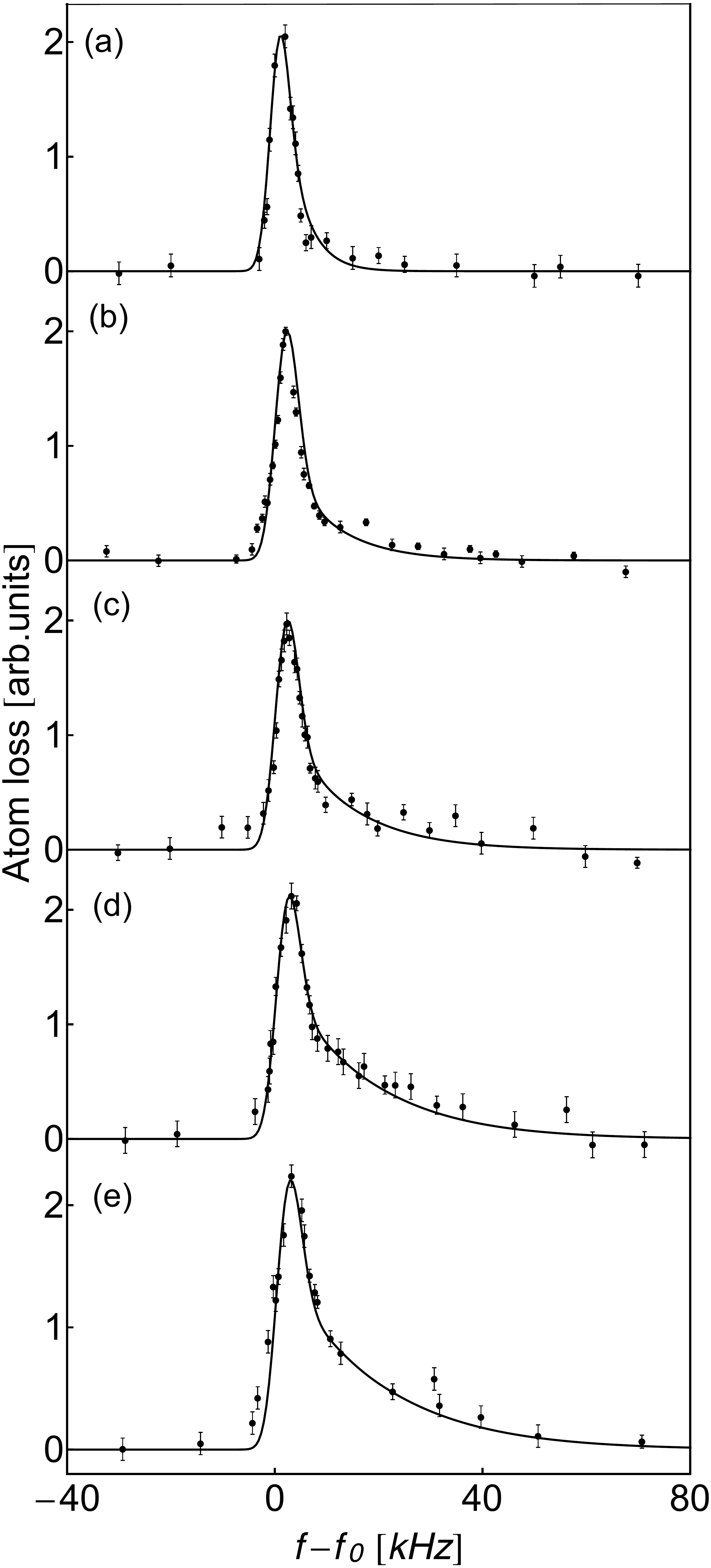} 
		\caption{ RF spectra of loss of atoms in lattice site 38 for different radial trap frequencies of (a) 1.5 kHz (b) 7.5 kHz (c) 9.5 kHz (d) 16.1 kHz and (e) 19.9 kHz, with $\omega_{ax}/2\pi = 260$ Hz, $\delta f = 100$ kHz. The solid lines are fits to the data points based on the self-consistent mean-field model for a BEC plus thermal cloud convoluted with a Gaussian magnetic noise function with FWHM = 4.3 kHz, as described in the text.}
                       \vspace{-1.5em}                
		\label{figure9}
	\end{center}
\end{figure}

\begin{figure}[htbp]
\centering  
\includegraphics[width=0.34\textwidth]{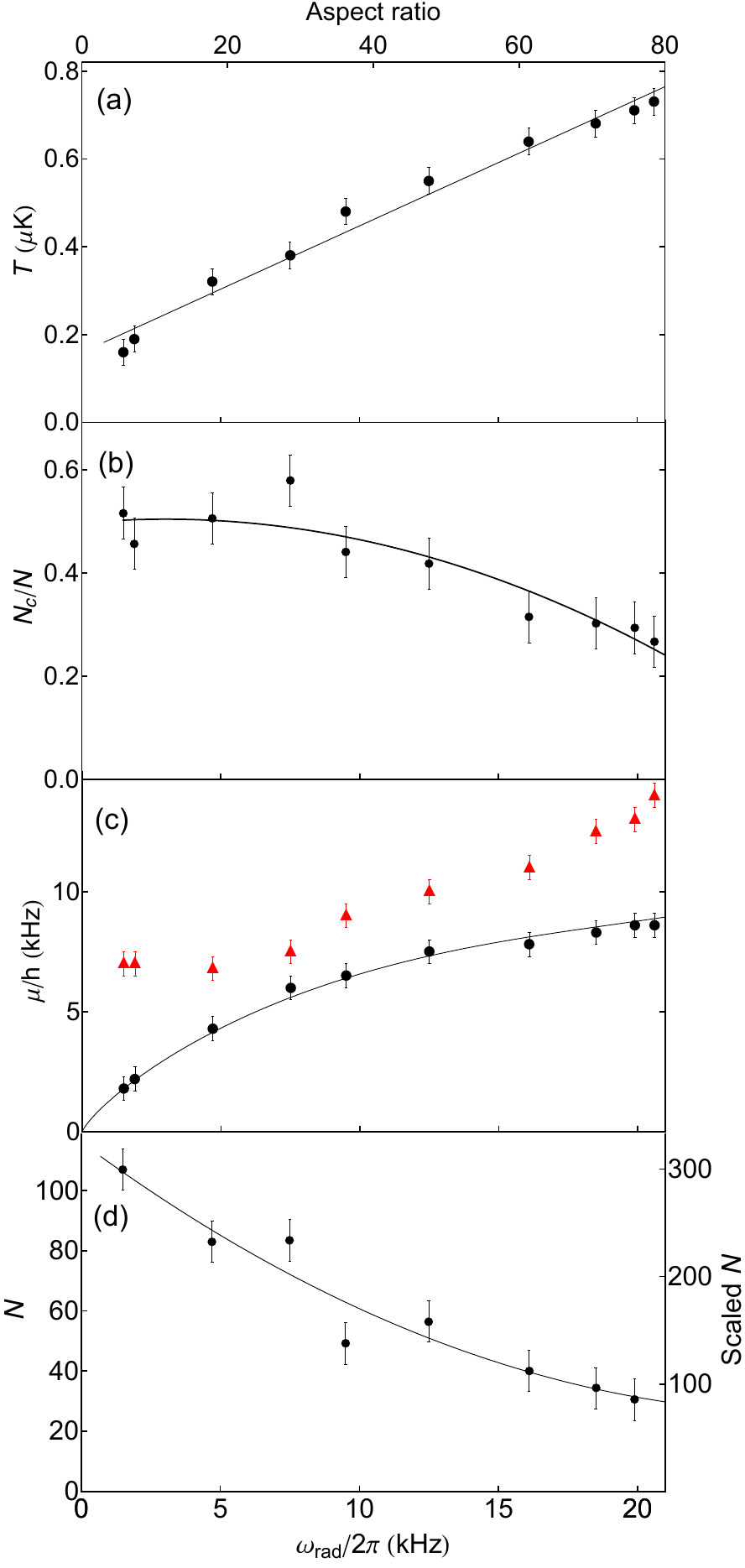} 
\caption[]{(a) Temperature (b) condensate fraction (c) chemical potential and (d) atom number per site versus radial trap frequency for lattice site 38, with $\omega_{ax}/2\pi = 260$ Hz, $\delta f = 100$ kHz. The black points in (a), (b), (c) were determined from fits to the RF spectra in Fig. \ref{figure9}, as described in the text. The red triangles in (c) were determined from fits to the RF spectra without allowing for magnetic noise broadening. The points in (d) were determined from absorption imaging measurements, where the atom number on the right vertical axis has been scaled by a factor of 2.8 (see text). The solid line in (c) represents a fit of $\omega_{rad}^{4/5} N(\omega_{rad})^{2/5}$ to the data points using the polynomial fit to the scaled $N(\omega_{rad})$ values in (d). The solid lines in (a), (b), (d) are polynomial ((b), (d)) or straight-line ((a)) fits to the data points.}
\label{figure10}
  \vspace{-0.8em}   
\end{figure}

It is of interest to investigate how the condensates survive as the trap frequency is increased to produce tighter lattice traps. Figure \ref{figure9} shows RF spectra recorded for lattice site 38 as the radial trap frequency is raised from $\omega_{rad}/2\pi$ = 1.5 kHz to 19.9 kHz, with the axial trap frequency kept fixed at $\omega_{ax}/2\pi$ = 260 Hz, and the trap depth	$\delta f=100$ kHz. The spectra exhibit a pronounced BEC component up to the highest trap frequencies investigated. The solid lines in Fig.~\ref{figure9} represent fits to the data obtained from the self-consistent mean-field model convoluted with a Gaussian magnetic noise function with FWHM = 4.3 kHz and with the atom number constrained to match the scaled atom number determined from the absorption imaging measurements (Fig. \ref{figure10}(d)). Satisfactory fits are obtained for all spectra over the range of trap frequencies investigated.

For very high radial trap frequencies ($\omega_{rad}>\mu/\hbar$) we expect a crossover from 3D to 1D behaviour. However, even for the highest radial trap frequencies the observed RF spectra exhibit clear bimodal profiles. Although it is not strictly valid in this regime, we used the 3D self-consistent mean-field model to fit the spectra in Fig. \ref{figure9} for all $\omega_{rad}$, including large $\omega_{rad}$, in order to estimate the spectral widths of the narrow (BEC-like) and broad (thermal) components. In particular, we have used this to provide an estimate for the temperature of the atomic gas through the crossover.

Figure \ref{figure10} shows the variation of temperature, condensate fraction and chemical potential determined from the fits together with the atom number per site determined from the absorption imaging measurements, over the range $\omega_{rad}/2\pi$ = 1.5 kHz to 20.6 kHz, corresponding to aspect ratios of 6 to 79. The temperatures determined from the fits increase approximately linearly from 0.16~$\mu$K at $\omega_{rad}/2\pi$ = 1.5 kHz to 0.73 $\mu$K at $\omega_{rad}/2\pi$~=~20.6 kHz. The increase in temperature is accompanied by a factor of about two reduction in condensate fraction. After allowing for magnetic noise broadening, the chemical potential closely follows the expected scaling for a 3D condensate $\mu =A\omega_{rad}^{4/5}N^{2/5}$	(solid line in Fig.~\ref{figure10}(c)), which is determined from $\mu = \frac{1}{2}\hbar\omega_{HO} \left(\frac{15Na_s}{a_{HO}}\right)^{2/5}$, where  $a_{HO} = \left( \frac{\hbar}{m\omega_{HO}}\right)^{1/2}$ and $\omega_{HO}=\bar{\omega}=(\omega_{rad}^2\omega_{ax})^{1/3}$, using the $N(\omega_{rad})$ values from the polynomial fit to the scaled atom numbers in Fig. \ref{figure10}(d). The pre-factor $A$ used to obtain the solid line in Fig. \ref{figure10}(c) is in reasonable agreement (17\% lower) with the theoretical factor $A = \frac{1}{2}\hbar\left(\frac{15a_sm^{1/2}\omega_{ax}}{\hbar^{1/2}}\right)^{2/5}$.

\begin{figure}[htbp]
	\begin{center}
		\includegraphics[width=0.45\textwidth]{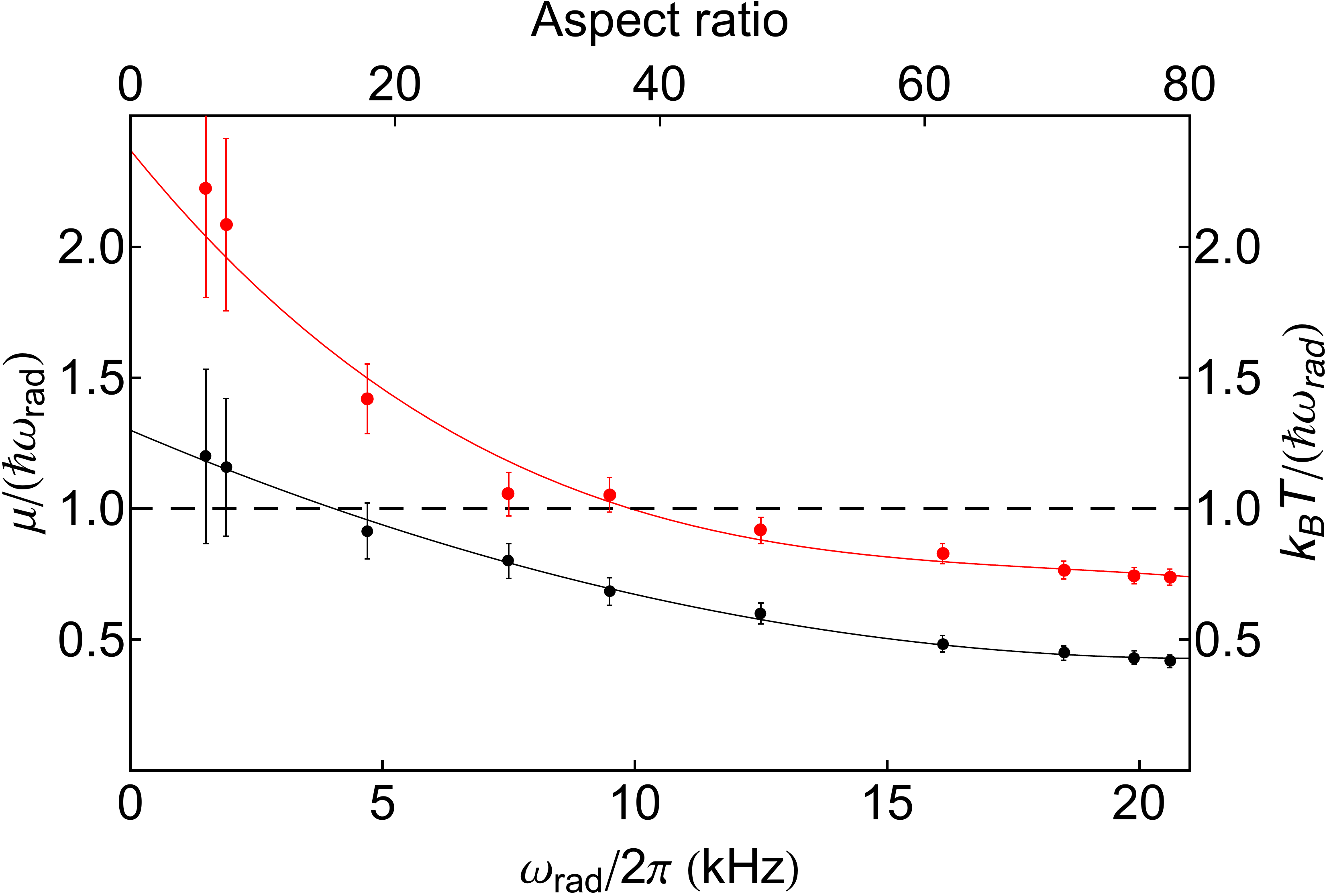} 
		\caption{ (Colour online) $\mu/(\hbar\omega_{rad})$ (black points) and $k_BT/(\hbar\omega_{rad})$ (red points) versus radial trap frequency for $\omega_{ax}/2\pi = 260$ Hz, $\delta f = 100$ kHz. The region where both black and red points lie below the dashed line indicates the quasi-1D Bose gas regime. The solid lines are polynomial fits to the data points. }
                       \vspace{-1.5em}                
		\label{figure11}
	\end{center}
\end{figure}

In Fig. \ref{figure11} we plot	$\mu/(\hbar\omega_{rad})$ and $k_BT/(\hbar\omega_{rad})$ against radial trap frequency. For $\omega_{rad}/2\pi > 10$ kHz, both the chemical potential $\mu$ and the thermal energy $k_BT$	lie below the energy of the lowest radial vibrational excited state $\hbar\omega_{rad}$, which represents the quasi-1D Bose gas regime \cite{Kinoshita06,mazets10,Jacqmin11, Armijo10, Buchoule11}. The fitted temperature is seen to increase approximately linearly through the crossover, and for the highest radial trap frequencies (20.6 kHz) the temperature is a factor of five larger than in the 3D regime. An explanation of the observed increase in temperature extracted from the 3D Hartree-Fock mean-field model could be reduced efficiency of evaporation due to suppression of rethermalising collisions in the quasi-1D regime. However, to conclusively determine the cause of this increase we require more sophisticated (beyond mean-field) models of the expected RF spectra in the crossover between the 3D and 1D regimes. 

A 1D Bose gas may be characterised by the Lieb-Liniger interaction parameter $\gamma\approx\frac{2a_s}{n_\mathrm{{1D}}l_{rad}^2}$ and the dimensionless temperature $t=\frac{2\hbar^2k_BT}{mg^2}$ \cite{Jacqmin11}. Here $n_\mathrm{{1D}}= N/l_{ax}$ is the linear atom density in the axial direction, $\l_{rad}= \sqrt{\hbar/(m\omega_{rad})}$ is the radial oscillator length (assumed to be $>>a_s$), $l_{ax}$ is the cloud rms length which is assumed to scale as the Thomas-Fermi radius $R_{ax} = \sqrt{2\mu/(m\omega_{ax}^2})$, and $g\approx 2\hbar\omega_{rad}a_s$ is the 1D coupling constant. From the above, $\gamma$ scales as $\omega_{rad}/(N\omega_{ax}^{11/15})$ and $t$ scales as $T/\omega_{rad}^2$. To reach the strongly interacting, strongly correlated 1D Tonks-Girardeau regime in a trapped Bose gas \cite{lieb63,lieb63-2} requires $\gamma >1$ and $t<1$ \cite{Jacqmin11}. For the highest trap frequency ($\omega_{rad}/2\pi = 20.6$ kHz) used in our magnetic lattice, the interaction parameter $\gamma\approx 0.3$ and $t\approx 70$. It may be possible in the future to reach the Tonks-Girardeau regime in our magnetic lattice using, for example, $\omega_{rad}/2\pi= 50$ kHz, $\omega_{ax}/2\pi = 100$ Hz, $N = 50$ and $T = 50$ nK, which give $\gamma\approx 2$ and $t\approx 0.9$.\vspace{0.2cm}

{\bf{D. RF spectra for various holding times}}\vspace{0.2cm}

\begin{figure}[htbp]
	\begin{center}
		\includegraphics[width=0.30\textwidth]{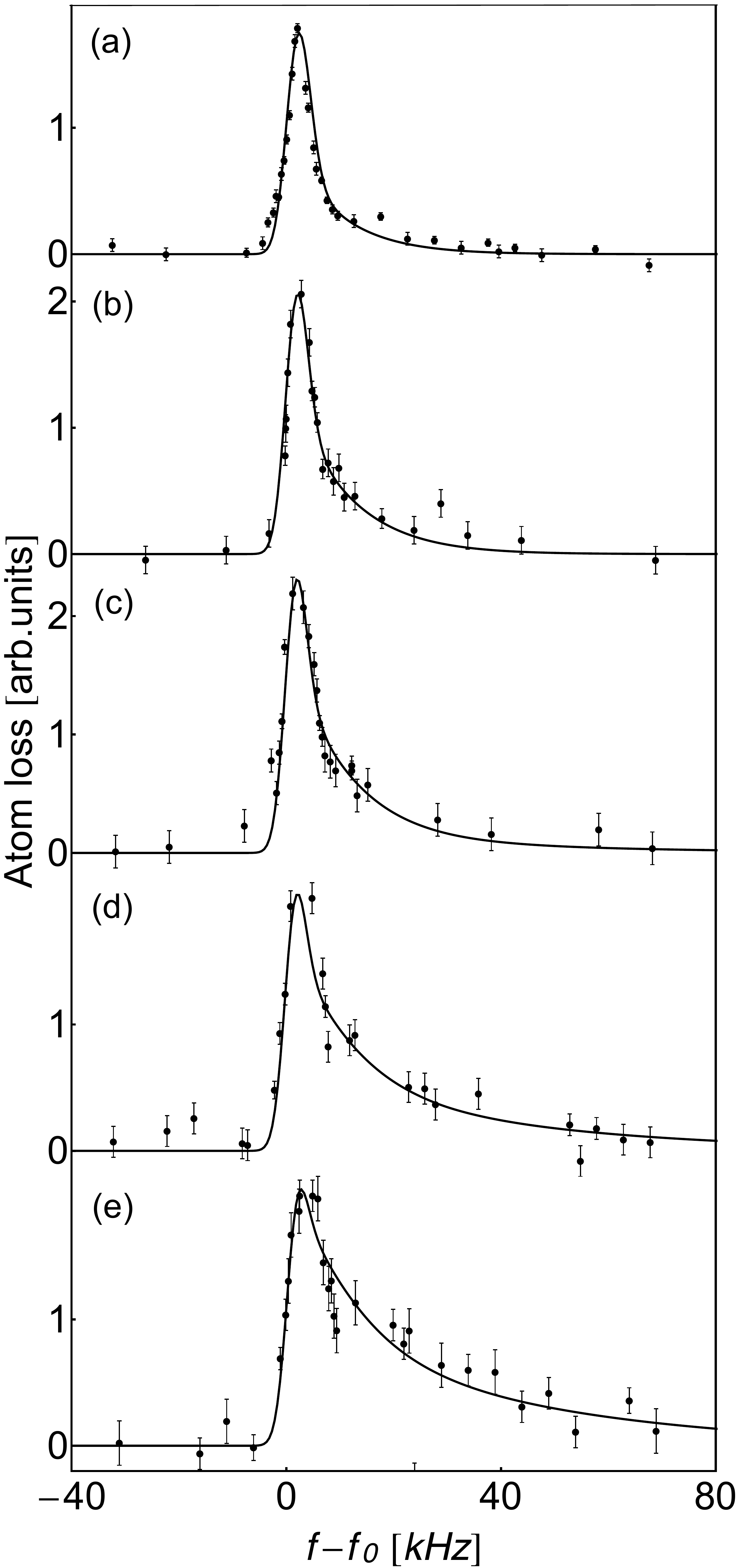} 
		\caption{ RF spectra of loss of atoms in lattice site 38 recorded for holding times of (a) 0 ms (b) 250 ms (c) 500 ms (d) 750 ms and (e) 1000 ms, after evaporative cooling to $\delta f= 100$ kHz, with $\omega_{rad}/2\pi = 7.5$ kHz, $\omega_{ax}/2\pi = 260$ Hz. The solid lines in (a) and (b) are fits to the data points based on the self-consistent mean-field model for a BEC plus thermal cloud convoluted with a Gaussian magnetic noise function with FWHM = 4.3 kHz, as described in the text. The solid lines in (c), (d) and (e) are fits based on a two-temperature model with $T_1 = 0.4$ $\mu$K for the BEC plus thermal cloud component and $T_2 = 1.5$ $\mu$K for the broad component, as described in the text. }
                       \vspace{-1.5em}                
		\label{figure12}
	\end{center}
\end{figure}

Figure \ref{figure12} shows RF spectra recorded for a range of holding times up to 1000 ms after evaporative cooling to $\delta f = 100$ kHz in the magnetic lattice, for lattice site 38 with $\omega_{rad}/2\pi = 7.5$ kHz, $\omega_{ax}/2\pi = 260$ Hz. For the spectra taken for $t_{hold} \ge 500$ ms we were unable to obtain satisfactory fits with the self-consistent mean-field model using a single temperature, nor could we obtain satisfactory fits with a Hartree-Fock-Bogoliubov finite-temperature model \cite{Hutchinson97} using a single temperature.

\begin{figure}[htbp]
\centering  
\includegraphics[width=0.35\textwidth]{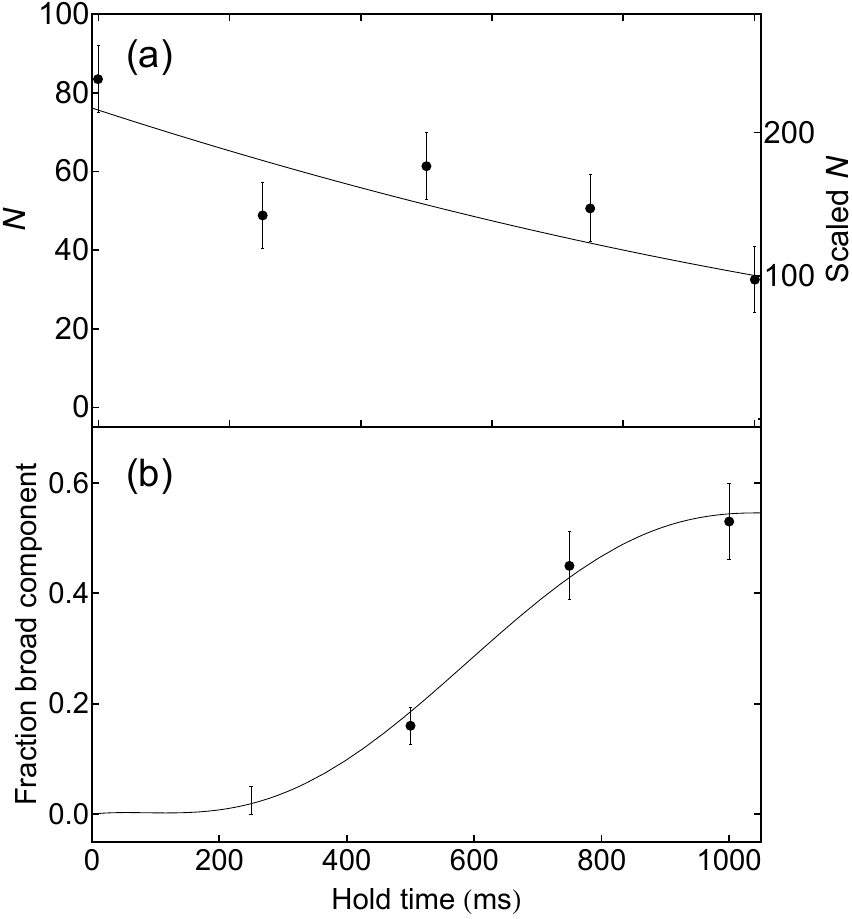} 
\caption[]{(a) Atom number determined from absorption imaging and (b) fraction of atoms in the broad $T_2= 1.5$ $\mu$K component in Fig. \ref{figure12} versus holding time for lattice site 38 with $\delta f = 100$ kHz, $\omega_{rad}/2\pi = 7.5$ kHz, $\omega_{ax}/2\pi = 260$ Hz. The error bar at the holding time of 250 ms represents an upper limit estimated from fits to the RF spectrum. The solid lines are polynomial fits to the data points. The atom number on the right vertical axis in (a) corresponds to the atom number scaled by a factor of 2.8 (see text).}
\label{figure13}
  \vspace{-0.8em}   
\end{figure}

To obtain reasonable fits, the spectra for long hold times in Fig. \ref{figure12} are fitted with a two-temperature model, in which we assume a fixed $T_1 = 0.4$ $\mu$K for the BEC plus thermal cloud component and $T_2 = 1.5$ $\mu$K for the broad component, convoluted with a Gaussian noise function with FWHM = 4.3 kHz and constrained so that the total atom number derived from the fit matches the scaled atom number determined from the absorption imaging measurements (Fig. \ref{figure13}(a)).

\vspace{-0.11em}The total atom number per site $N$ for the condensate plus thermal cloud determined by absorption imaging (Fig. \ref{figure13}(a)) decays with a half-life $t_{1/2}\sim 0.9 \pm 0.3$~s, which is consistent with the three-body decay half-life for a pure condensate, $t_{1/2}\sim 0.6$ $\pm$ 0.2, estimated using $N_c(t)=N_c(0)[1+\alpha N_c(0)t]^{-5/4}$, where  $\alpha=L_3\frac{(m\bar{\omega}/\hbar)^{12/5}}{14\times 15^{1/5}\pi^2a_s^{6/5}}$ \cite{Dennis12}, $L_3=(5.8\pm1.9)\times10^{-30}$ cm$^6$s$^{-1}$ \cite{Burt97}, $a_s = 5.3$ nm and $\bar{\omega}/2\pi = 2.40$ kHz. The above half-life, $t_{1/2}\sim 0.9 \pm 0.3$ s, is also similar (within about 10\%) to the three-body decay half-life for an ultracold thermal cloud in which the six-times larger $L_3$ coefficient \cite{Burt97} is approximately compensated by the much smaller peak atom density for the thermal cloud (Fig. \ref{figure2}(a)). We attribute the observed decay of the atom number with holding time mainly to three-body loss.

The decay of the total atom number $N$ with holding time (Fig. \ref{figure13}(a)) is accompanied by a growth in the fraction of atoms in the broad component (Fig. \ref{figure13}(b)). Previous work (e.g., \cite{Cornell99}) has shown that an ultracold rubidium cloud held in a magnetic trap for several seconds after evaporative cooling can develop ``wings" in the density spatial profile that cannot be fitted with a single Gaussian. This has been interpreted in terms of the outer region of the magnetic trap becoming filled with a dilute, high energy halo of trapped atoms, the ``Oort cloud'', which can remain trapped in a deep magnetic trap without being in thermal equilibrium with the ultracold cloud. The high energy cloud may be produced by a number of mechanisms \cite{Cornell99} such as energetic inelastic decay products of three-body recombination. Due to its very low density, the high energy cloud is only weakly coupled to the ultracold atom cloud, with occasional collisions transferring energy between the two trapped components. In the present experiment we attribute the appearance of the broad high temperature component after hold times $>$ 250 ms to rubidium atoms that have acquired additional energy through collisions with energetic three-body decay products and remain trapped.  The observed delayed onset of the broad high temperature component (Fig.~\ref{figure13}(b)) is interpreted as due to the time required for the high energy cloud to accumulate in the traps and to transfer energy to part of the ultracold atom cloud.\vspace{-0.4cm}
\section{V.	SUMMARY AND OUTLOOK}\vspace{-0.4cm}
We have investigated Bose-Einstein condensates of $^{87}$Rb $|F=1,m_F=-1>$ atoms trapped in multiple sites of a one-dimensional 10 $\mu$m-period magnetic lattice using site-resolved RF spectroscopy for various sites across the lattice and for a range of trap depths, trap frequencies and holding times in the lattice. The site-to-site variation of the trap bottom, temperature, condensate fraction and chemical potential across the magnetic lattice indicates that the magnetic lattice is remarkably uniform. In particular, the trap bottoms, which could be accurately determined, show site-to-site variations of only $\pm$ 0.4 mG.

At the lowest radial trap frequency (1.5 kHz), temperatures down to 0.16 $\mu$K were achieved in the magnetic lattice and at the smallest trap depth (50 kHz) condensate fractions up to 80\% were observed. With increasing radial trap frequency (up to 20 kHz) large condensate fractions continued to persist and the highly elongated atom clouds approached the quasi-1D Bose gas regime. The temperature estimated from analysis of the spectra was found to increase by a factor of about five which may be due to suppression of rethermalising collisions in the quasi-1D Bose gas. It would be of interest to extend the operating conditions of the present magnetic lattice to enable the elongated atom clouds to enter the strongly interacting Tonks-Girardeau regime, in which the atom clouds behave like a gas of non-interacting 1D fermions.

Measurements taken for holding times up to 1000 ms in the magnetic lattice indicated an atom number decay with a half-life of about 0.9 s due to three-body recombination and the appearance of a broad high temperature ($\sim 1.5 $ $\mu$K) component. The broad component is attributed to atoms that have acquired energy through collisions with energetic three-body decay products and remain trapped.

In the present 10 $\mu$m-period magnetic lattice the multiple BECs represent an array of isolated condensates with no tunnelling or interaction between the atoms in neighbouring lattice sites and no phase coherence between the condensates. Such a magnetic lattice seems suited for conducting experiments on Rydberg-interacting quantum systems which exploit the long-range dipolar interaction between atoms excited to Rydberg states \cite{Leung14,Saffman10}. In a 10 $\mu$m-period 2D lattice, the interaction-driven level shift between, e.g., two $n \approx$ 80s Rydberg atoms separated by 10 $\mu$m, is still several MHz, which is much faster than the decay rate of Rydberg states \cite{Saffman10}, while the fragile Rydberg atoms are trapped about 5 $\mu$m from the chip surface which should be sufficient to minimize surface effects \cite{Leung14, Tauschinsky10}.

We have recently fabricated magnetic microstructures with periods down to 0.7 $\mu$m to create 1D and 2D square and triangular magnetic lattices in which the tunnelling times between neighbouring lattice sites are estimated to be of the order of tens of milliseconds \cite{ivan14}. To perform quantum tunneling experiments in a 0.7 $\mu$m magnetic lattice a number of challenges need to be considered. First, high quality magnetic potentials that are smooth and highly periodic (to better than 1\%) are needed in order to minimize effects due to disorder and any fragmentation of the elongated atom clouds. Secondly, the atoms will be trapped at distances of typically 0.35 $\mu$m from the chip surface, and therefore surface effects such as attractive van der Waals forces and thermally induced spin-flips produced by thermal currents in the nearby conducting film need to be considered \cite{Leung11,ivan14}. Thirdly, in order to operate at the barrier heights required for quantum tunnelling experiments (e.g., $V_0 = 12E_r = 20$ mG for $^{87}$Rb $F=1$ atoms, where $E_r=\frac{\pi^2\hbar^2}{2ma^2}$), stray magnetic fields and magnetic noise need to be compensated to better than 1 mG. Finally, due to the tight confinement in the 0.7 $\mu$m-period magnetic lattices atomic states with low inelastic collision rates, such as fermionic atoms, need to be chosen.\vspace{-0.8cm}
\section{ACKNOWLEDGMENTS}\vspace{-0.35cm}
We thank Mandip Singh, Brenton Hall, Chris Vale and Karen Kheruntsyan for fruitful discussions and James Wang for assistance with the fabrication and characterisation of the magnetic microstructure. The project is funded by an Australian Research Council Discovery Project grant (Grant No. DP130101160).



\vspace{0.3cm}
\begin{thebibliography}{30}
\bibitem{Ghanbari06} S. Ghanbari, T. D. Kieu, A. Sidorov, and P. Hannaford, J. Phys. B {\bf 39}, 847,  (2006).
\bibitem{Gerritsma07}R. Gerritsma {\it et al.}, Phys. Rev. A {\bf 76}, 033408 (2007).
\bibitem{Singh08} M. Singh {\it et al.}, J. Phys. B. {\bf 41}, 065301 (2008).
\bibitem{Whitlock09}S. Whitlock, R. Gerritsma, T. Fernholz, and R. J. C. Spreeuw, New J. Phys. {\bf 11}, 023021 (2009).
\bibitem{Abdelrahman10} A. Abdelrahman, M. Vasiliev, K. Alameh, and P. Hannaford, Phys. Rev. A {\bf 82}, 012320 (2010).
\bibitem{Schmied10}R. Schmied, D. Leibfried, R. J. C. Spreeuw, and S. Whitlock, New J. Phys. {\bf 12}, 103029 (2010).
\bibitem{Leung11} V. Y. F. Leung, A. Tauschinsky, N. J. van Druten, and R. J. C. Spreeuw, Quant. Inf. Process. {\bf 10}, 955 (2011).
\bibitem{jose14} S. Jose, P. Surendran, Y. Wang, I. Herrera, L. Krzemien, S. Whitlock, R. McLean, A. Sidorov, and P. Hannaford, Phys. Rev. A {\bf 89}, 051602(R) (2014).
\bibitem{Leung14} V. Y. F. Leung, D. R. M. Pijn, H. Schlatter, L. Torralbo-Campo, A. L. La Rooij, G. B. Mulder, J. Naber, M. L. Soudijn, A. Tauschinsky, C. Abarbanel, B. Hadad, E. Golan, R. Folman, and R. J. C. Spreeuw, Rev. Sci. Instrum. {\bf 85}, 053102 (2014).
\bibitem{ivan14} I. Herrera, Y. Wang, P. Michaux, P. Surendran, S. Juodkazis, S. Whitlock, R. J. McLean, A. Sidorov, D. Nissen, M. Albrecht and P. Hannaford, arXiv:1410.0528 (2014).
\bibitem{Burt97} E. Burt {\it et al.}, Phys. Rev. Lett. {\bf 79}, 337 (1997).
\bibitem{Soeding99} J. Soeding {\it et al.}, Appl. Phys. B {\bf 69}, 257 (1997).
\bibitem{Kinoshita06} T. Kinoshita, T. Wenger, and D. S. Weiss, Nature {\bf 440}, 900 (2006).
\bibitem{mazets10} I. E. Mazets and J. Schmiedmayer, New J. Phys {\bf 12}, 055023 (2010).
\bibitem{Jacqmin11} T. Jacqmin, J. Armijo, T. Berrada, K. V. Kheruntsyan, and I. Bouchoule, Phys. Rev. Lett. {\bf 106}, 230405 (2011).
\bibitem{Gerbier04}F. Gerbier {\it et al.}, Phys. Rev. A {\bf 70}, 013607 (2004).
\bibitem{Wang05}J. Y. Wang {\it et al.}, J. Phys. D {\bf 38}, 4015 (2005).
\bibitem{SJose13}S. Jose, PhD Thesis, Swinburne University (2013).
\bibitem{radia} http://www.esrf.fr/Accelerators/Groups/InsertionDevices/ Software/Radia.
\bibitem{hughes97}I. G. Hughes, P. A. Barton, T. M. Roach, and E. A. Hinds, J. Phys. B  {\bf 30}, 2119 (1997).
\bibitem{Jauregui01}R. J{\'a}uregui, Phys. Rev. A  {\bf 64}, 053408 (2001).
\bibitem{Armijo10}J. Armijo, T. Jacqmin, K. V. Kheruntsyan, and I. Bouchoule, Phys. Rev. Lett. {\bf105}, 230402 (2010).
\bibitem{Buchoule11} I. Bouchoule, N. J. Van Druten and C. J. Westbrook, In Atom Chips, Ch. 11, p 331, Eds. J. Reichel and V. Vuletic, Wiley-VCH, Verlag  (2011).
\bibitem{lieb63}E. H. Lieb and W. Linager, Phys. Rev. {\bf130}, 1605 (1963).
\bibitem{lieb63-2}E. H. Lieb, Phys. Rev. {\bf130}, 1616 (1963).
\bibitem{Hutchinson97}D. A. W. Hutchinson, E. Zaremba, and A. Griffin, Phys. Rev. Lett. {\bf 78}, 1842 (1997).
\bibitem{Dennis12}G. R. Dennis, M. J. Davis and J. J. Hope, Phys. Rev. A {\bf 86}, 013640 (2012).
\bibitem{Cornell99}E. A. Cornell, J. R. Ensher, and C. E. Wieman, In Proceedings of the International School of Physics,``Enrico Fermi$"$, Course CXL, Eds. M. Inguscio, S. Stringari, and C. E. Wieman (IOS Press, Amsterdam) 1999, p 15.
\bibitem{Saffman10} M. Saffman, T. G. Walker, and K. M$\o$lmer, Rev. Mod. Phys. {\bf 82}, 2313 (2010).
\bibitem{Tauschinsky10}A. Tauschinsky, R. M. T. Thijssen, S. Whitlock, H. B. van Linden van den Heuvell, and
R. J. C. Spreeuw, Phys. Rev. A {\bf 81}, 063411 (2010).



\end{thebibliography}
\end{document}